\documentclass[a4paper,12pt]{article}
 \pdfoutput=1
\usepackage[pdftex]{graphicx}
\usepackage{subfigure}

\newcommand{\be}{\begin{equation}}
\newcommand{\ee}{\end{equation}}

\newcommand{\bea}{\begin{eqnarray}}
\newcommand{\eea}{\end{eqnarray}}

\newcommand{\p}{\partial}

\newcommand{\nn}{\nonumber \\}
\newcommand{\f}{\frac}
\newcommand{\w}{\wedge}
\newcommand{\ra}{\rightarrow}
\begin{document}
\thispagestyle{empty}
\begin{flushright}
{\bf arXiv:1209.3559}
\end{flushright}
\begin{center} \noindent \Large \bf Fermi-like Liquid 
From 
Einstein-DBI-Dilaton System

\end{center}

\bigskip\bigskip\bigskip
\vskip 0.5cm
\begin{center}
{ \normalsize \bf   Shesansu Sekhar Pal
}

%\vskip 0.5cm

\vskip 0.5 cm
%Saha Institute of Nuclear Physics,
%Sector 1, Block AF,  700064,  Kolkata, India

Centre of Excellence in Theoretical and Mathematical Sciences, SOA University,  Bhubaneswar, 751030, India
\vskip 0.5 cm
\sf { shesansu${\frame{\shortstack{AT}}}$gmail.com 
}
\end{center}
\centerline{\bf \small Abstract}

We have obtained an  expression of the entropy density depending on the scale transformation of the spatial directions in the field theory. It takes the following form in $d+1$ dimensional bulk spacetime: $s\sim T^{\f{\delta(d-1)-\theta}{z}}_H$, where $z$ and $T_H$ are the dynamical exponent and temperature in the field theory, respectively. $\theta$ is related to the scaling violation exponent, whereas $\delta$ gives us the information about the scaling behavior of the spatial field theoretic direction. This we demonstrate by finding solutions to the  Einstein-DBI-dilaton system in generic spacetime dimensions. Upon restricting to $d=3$, we show  the linear temperature dependence of the  specific heat and  inverse quadratic  temperature dependence of the  resistivity for $z=2,\theta=0$ and $\delta=1$, which resembles that of the Fermi-like liquid.
Whereas for  $z=2,~\theta=-2$ and $\delta=0$ gives us a   solution that is conformal to $AdS_2\otimes R^2$, which  resembles with the non-Fermi-like liquid. Moreover, it shows the logarithmic violation of the entanglement entropy when the entangling region is of the strip type.

\newpage

\section{Introduction}
There has been a lot of activity in trying to understand the scale invariant gravitational solution, which  asymptotes to AdS at UV and at IR, it can behave either like $AdS_2$ or Lifshitz type.  Recently, a non-scale invariant gravitational solution found in \cite{Charmousis:2010zz} and \cite{Gouteraux:2011ce} has been interpreted in \cite{Huijse:2011ef} to give  the compressible state of the matter which exhibits the hidden Fermi surfaces, using holography \cite{Maldacena:1997re}.  
In this context, it is suggested that
 the field theory directions and the  invariant interval of the bulk scale in the following way
\be\label{scale_symm}
 t\ra \lambda^{z} t,\quad x_i \ra \lambda^{\delta} x_i,\quad r\ra \f{r}{\lambda} ,\quad ds \ra \lambda^{\gamma} ds,\quad %{\rm for}~~{\delta}=1
\ee 
where $z$ is the dynamical exponent, $\gamma$ is the scaling violation exponent, which is related to the hyperscaling violation exponent as in \cite{Huijse:2011ef}, and more importantly, the spatial directions scale linearly, i.e., ${\delta}=1$. Note that $\delta$ can take only two values\footnote{
See Appendix A for further explanation of eq(\ref{scale_symm}) and  references \cite{Dong:2012se}-
\cite{Alishahiha:2012qu}  for further studies.}, namely, $\delta=0,~1$.

 In this paper, we shall construct explicit  solutions with vanishing $\gamma,~\delta$ and non-vanishing $\gamma,~\delta$  and study its consequences.  We construct such  bulk solutions with the help of gravity, U(1) gauge field and a scalar field. 
In order to do so, we have considered a space filling brane, whose action is described by  the Dirac-Born-Infeld (DBI) action. Upon considering the back reaction of the DBI action and that of the scalar field  on to the geometry in $d+1$ dimensional bulk spacetime, makes   the metric looks as
\be\label{most_general_geometry_with_all_possibilities}
ds^2_{d+1}=r^{-2\gamma}[-r^{2z}dt^2+ r^{2\delta}dx^idx^j\delta_{ij}+\f{dr^2}{r^2 }]\equiv r^2 ds^2_L.
\ee

It is easy to notice that the metric can be written as a spacetime which is  conformal to the Lifshitz spacetime \cite{Kachru:2008yh}. In which case,  the geometry, $ds^2_L$, scales as \cite{Taylor:2008tg}, \cite{Pal:2009yp},\cite{Kachru:2011ps} and \cite{Inbasekar:2012sh}  
\be
t\ra \lambda^{z} t,\quad x_i\ra \lambda^{\delta} x_i,\quad r\ra \f{r}{\lambda},\quad {\rm for}\quad \delta=0,~1.
\ee
For $\delta=0$,   this particular  \lq\lq{}Lifshitz spacetime\rq\rq{} can be re-written  as an $AdS_2\times R^2$. In fact, for this choice of, $\delta$,   the geometry $ds^2_L$ is not in the sense of \cite{Kachru:2008yh} because of the scaling behavior of the spatial directions\footnote{ Moreover, the Lifshitz spacetime as defined in   \cite{Kachru:2008yh} suffers from the null curvature singularity and are unstable  \cite{Horowitz:2011gh}.}.
For an earlier study of the Einstein-Maxwell-dilaton system see e.g.,  
\cite{Charmousis:2010zz}, \cite{Gouteraux:2011ce}  \cite{Gubser:2009qt},
 \cite{Goldstein:2009cv} and 
\cite{Chen:2010kn}.

Let us recall from \cite{Huijse:2011ef}, a theory which exhibits the scaling violation exponent for  $\delta=1$ 
 should see a reduced entropy. In fact, for $d-1$ number of spatial directions with $\gamma=\theta/(d-1)=(d-2)/(d-1)$, the entropy should go as $s\sim T^{\f{d-1-\theta}{z}}_H$, where $T_H$ is the Hawking temperature.
However, for ${\delta}=0$ and $\gamma=-1$, there do not arises any change in the entropy.
In fact, the entropy behaves like that of a scale invariant solution\footnote{For this choice of $\delta$, with the boundary at $r=\infty$, implies $\gamma <0$. }. 
It is due to the fact that  in $d+1$ dimensional spacetime the   
 complete solution  can be written as a solution that is conformal to $AdS_2\times R^{d-1}$ solution and  reads as
 \be
 ds^2=R^2 r^{-2\gamma/z}\left[-r^2 f(r)dt^2+dx^idx^j\delta_{ij}+\f{dr^2}{r^2f(r)} \right],\quad f(r)=1-(r_h/r)^{\f{z+(1-\gamma)(d-1)}{z}}.
 \ee
 
It is easy to see that  the   entropy density  goes as $s\sim T^{\f{-\gamma(d-1)}{z}}_H$. The temperature dependence of the entropy for these two cases in $d+1$ dimensional spacetime  can be summarized as follows:
 \be
  s \sim  \left\{ 
  \begin{array}{l l}
   T^{\f{d-1-\theta}{z}}_H  & \quad \rm{for} \quad \delta=1 \\ 
  T^{\f{-\gamma(d-1)}{z}}_H  & \quad \rm{for}\quad \delta=0\\
  \end{array}
 \right.
\ee
Upon combining the formulas together in  $d+1$ dimensional bulk spacetime dimensions, the formula for the entropy density with the scaling violation exponent, $\theta\equiv\gamma(d-1)$, and the spatial scaling dimension, $\delta$, can be written  as 
 \be\label{generic_entropy}
 s\sim T^{\f{\delta(d-1)-\theta}{z}}_H.
 \ee
We see there exists a non-zero entropy density for $\delta=1$ case in the limit of  $z\ra \infty$ for finite $\theta$ in the zero  temperature limit. However, for $\delta=0$ case, we get  the entropy density as   $s\sim T^{-\theta/z}_H$. 
 Note that this expression of the entropy density, eq(\ref{generic_entropy}), vanishes  in the limit of $z\ra \infty$ and $\theta\ra \infty$ by keeping the ratio $(-\theta/z)$ fixed \cite{Hartnoll:2012wm} at low temperature\footnote{ 
 More on  the expression of the entropy is discussed in   section 4.}.
 
 The specific heat in this case becomes
 \be
 c_v=T_H\left(\f{\p s}{\p T_H} \right)_V\sim \left[\f{\delta(d-1)-\theta}{z}\right]T^{\f{\delta(d-1)-\theta}{z}}_H.
 \ee
 Note that for $\delta=0$, in order to have positive specific heat, which is required for stability, should have negative $\theta/z$.
 
 Let us restrict the dimensionality of the spacetime to $3+1$. In this case,
the solution  is conformal to the  $AdS_2\times R^2$ solution,   see eq(\ref{bh_dbi_3+1_d_IR}),  and  has $\gamma=-1$ and $\delta=0$. 
For this case, the specific heat and the longitudinal conductivity   goes as
\be\label{nfl}
c_v\sim T^{2/z}_H,\quad \sigma\sim T^{-2/z}_H.
\ee

For $z=2$, the conductivity resembles with the non-Fermi liquid. 
Moreover, for such a choice of $\gamma$ and $\delta$, we do see the logarithmic violation of the entanglement entropy when the entangling region is of the strip type \cite{Ryu:2006bv}.
 
Moving on to the $\gamma=0$ and $\delta=1$ case, we  see the existence of a Fermi-like liquid, which follows  by doing an explicit computation of the transport and the thermodynamic quantity: the longitudinal conductivity as well as the specific heat. We found that, depending on the choice of $z$, the specific heat and  the longitudinal conductivity can have the linear and the inverse quadratic dependence on the temperature, respectively. This we demonstrate by finding an exact black hole solution to a $3+1$ dimensional Einstein-DBI-dilaton system. In which case, the spacetime asymptotes to a Lifshitz spacetime
with a non-trivial profile to the scalar field 
\be
ds^2=-r^{2z}f(r)dt^2+r^2(dx^2+dy^2)+
\f{dr^2}{r^2f(r)},\quad \phi(r)\sim log~r,\quad f(r)=1-(r_h/r)^{z+2}
\ee
and for some non-trivial form of the U(1) gauge field field strength such that it vanishes at IR in the zero temperature limit. In this case, the specific heat and the conductivity takes the following form
\be\label{fl}
c_V\sim T^{2/z},\quad \sigma
\sim T^{-4/z}.
\ee
The Fermi-like liquid  behavior follows when the dynamical exponent takes a specific value, $z=2$. More importantly, the entropy in the zero temperature limit vanishes (for finite $z$), suggesting the
compressible nature of the configuration. In this case, the computation of the  entanglement entropy for a strip does not show up the necessary logarithmic   term, as expected from the result of \cite{Huijse:2011ef}. 
% This finding in some sense supports the result of \cite{Ogawa:2011bz}. 
Further studies related to the Fermi liquid or the presence of Fermi surfaces are reported e.g.,  in \cite{Lee:2008xf}-\cite{Anantua:2012nj}. 

%It is interesting to note that for a specific value of the dynamically exponent, $z=2$, we see from eq(\ref{nfl}) and eq(\ref{fl})   that it\rq{}s only the conductivity that go from the non-Fermi liquid (NFL) type to the Fermi liquid (FL) type. 
%This happen  for the choice of the parameters as defined in eq(\ref{def_alpha_beta}).
 %This kind of ``jump" from NFL phase to FL phase is noted previously in \cite{Pal:2012gr}.

In the absence of the scalar field, we find an electrically charged black hole solution   in arbitrary spacetime  dimension at  UV whose form precisely matches with that of the solution found for the Born-Infeld black holes in \cite{Dey:2004yt} but   not the dyonic black hole solution in $3+1$ dimensional spacetime.
For these type of black hole solutions there exits a non-zero entropy even at zero temperature. This particular property is similar in nature  to that of the Reissner-Nordstrom (RN) black hole.

 The findings of the  paper  for the Einstein-DBI-dilaton  system in $d+1$ dimensional spacetime  are summarized in Table (\ref{tab_1}).

\begin{center}
\begin{table}\label{tab_1}
\begin{tabular}{ | l || l | l | l |}
    \hline
  Solutions at IR & Solutions at UV\\ \hline
For  $g_{MN}\neq 0,~F_{MN}\neq 0$ and $\phi = 0$ & For $g_{MN}\neq 0,~F_{MN}\neq 0$ and $\phi = 0$, eq(\ref{sol_UV_electrically_d+1})\\ 
& (a) generates charged AdS black hole solution \\
& in any arbitrary spacetime   dimensions;\\  
&  the entropy density in the vanishing \\& temperature limit remains non-zero,\\
&$s=\f{2\pi}{\kappa^2} T_b \f{\rho}{\sqrt{4\Lambda^2-T^2_b}},\quad {\rm as}~T_H \ra 0$;\\
&The chemical potential is not a continuous\\ &function of the charge density.\\
& No Log structure in the \\
&entanglement entropy.\\
\cline{2-2}
it generates $AdS_2\times R^{d-1}$, eq(\ref{ads2Rd-1}) & (b) Dyonic AdS black hole solution in $3+1$\\
&  dimensional spacetime, eq(\ref{dyonic_dbi}); \\ 
&  the entropy density  in the vanishing \\
Does not show log structure in the  & temperature limit remains non-zero,\\
entanglement entropy  for $d=2$ \cite{Swingle:2009wc}.& $s=\f{2\pi}{\kappa^2} T_b \f{\sqrt{\rho^2+\lambda^2 B^2}}{\sqrt{4\Lambda^2-T^2_b}},\quad {\rm as}~T_H \ra 0.$\\
& No Log structure in the \\
&entanglement entropy.\\
\hline\hline
For  $g_{MN}\neq 0,~F_{MN}\neq 0$ and $\phi \neq 0$, &For  $g_{MN}\neq 0,~F_{MN}\neq 0$ and $\phi \neq 0$, eq(\ref{geometry_dilaton_dbi}) \\
(a) Lifshitz   solution eq(\ref{lifshitz_sol}) with $\alpha\neq 0$,& Generates electrically charged black holes \\ 
$\beta\neq 0$ in $3+1$ dimensional spacetime;  & in generic $d+1$ dimensional spacetime.\\
the entropy density vanishes as & The specific heat can become negative \\  temperature vanishes, $s\sim T^{2/z}_H$; & for $\delta_1 >1/2$.\\
The specific heat, $c_V\sim T^{2/z}_H$;&\\
The longitudinal conductivity, $\sigma \sim T^{-4/z}_H$; &\\
No Log structure in the &\\entanglement entropy.&\\
\cline{1-1}
(b) Hyper scaling violating solution eq(\ref{bh_dbi_3+1_d_IR}) & \\
 but without decreasing the entropy with&\\
  $\alpha\neq 0,~\beta = 0$ in $3+1$ dimensional &\\spacetime;  
the entropy density vanishes as &\\  temperature vanishes, $s\sim T^{2/z}_H$; &\\
The specific heat, $c_v \sim T^{2/z}_H$; &\\
The longitudinal conductivity, $\sigma \sim T^{-2/z}_H$; &\\
Shows Log structure in the&\\ entanglement entropy.&\\
    \hline
  \end{tabular}
\caption{The summary of the solution of the Einstein-DBI-Dilaton system with two parameters $\alpha$ and $\beta$ and the potential, $V(\phi)$, as  defined in eq(\ref{def_alpha_beta}) and eq(\ref{potential}). The entanglement entropy is obtained for a strip.}
\end{table}
\end{center}

The paper is organized as follows. In section 2, we write down the effective action and its equation of motion. In section 3, we shall present the solution   at IR. In particular, the solution that shows the (non)Fermi-like liquid  behavior. In section 4, we generalize the  solutions at IR to arbitrary spacetime dimensions. 
In section 5, we find both the  electrically charged black hole and a dyonic solution for trivial and non-trivial  scalar field. 
%Then show that the electrically charged black hole solution found for the DBI actions are same as  that found for Born-Infeld actions.
In section 6 and 7, we compute the conductivity as well as the entanglement entropy, respectively.
And finally, we conclude in section 8. Some of the details are relegated to the Appendices.

\section{The action}

The action that we consider contains metric, the abelian gauge field and the scalar field as the degrees of freedom. In particular, the action involving the gauge field is the non-linear generalization of the Maxwell action, namely the Dirac-Born-Infeld action. The exact form of the action is\footnote{See \cite{Jarvinen:2011qe}, where the authors used a related action to study the holographic QCD in the Veneziano limit.}
\be\label{eh_dbi_dilaton}
S=\f{1}{2\kappa^2}\int d^{d+1} x\bigg[ \sqrt{-g}\bigg(R-2\Lambda-\f{1}{2}\p_M\phi\p^M\phi-V(\phi)\bigg)-T_bZ_1(\phi)\sqrt{-det\bigg([g] Z_2(\phi)+\lambda F\bigg)_{ab}}\Bigg],
\ee 
where $[g]_{ab}=\p_aX^M\p_b X^N g_{MN}$ is the induced metric on to the world volume of the brane. $T_b$  and $\Lambda$ are  the tension of the brane, and cosmological constant, respectively.  $F=dA$ is the two-form field strength. 
%In what follows, we shall consider the massless limit in which  the embedding fields are identified with the spacetime coordinates. 
Since, we are considering the brane to fill the entire space, 
 means $[g]_{ab}=g_{ab}$. The action as written down in  eq(\ref{eh_dbi_dilaton}) is in Einstein frame. The constant $\lambda$ is a dimension full object and has the dimension length$^2$ and in string theory it is identified with $\lambda=2\pi l^2_s$, where $l_s$ is the string length \cite{Myers:1999ps}. 
It is there to make the determinant dimensionless. The indices $a,~b$ are the world volume ones whereas the $M,~N$ etc denote the spacetime ones.  In the present case both kind of indices can take $d+1$ values.
Note that for small values of field strength, one can Taylor expand the determinant and obtain the Maxwell action\footnote{The term $det(gZ_2+\lambda F)$ can be expanded as $det(gZ_2)+det(\lambda F)+\f{1}{(d+1)!} \epsilon^{a_1\cdots a_{d+1}}\epsilon^{b_1\cdots b_{d+1}}(gZ_2)_{a_1b_1}\cdots (gZ_2)_{a_{d-1}b_{d-1}}F_{a_{d}b_{d}}F_{a_{d+1}b_{d+1}}+\cdots$. The ellipses denote higher even powers of $F$, which can be ignored in the dilute regime.  So, it is the quadratic in $F$ that gives the Maxwell action with a dilaton dependent Yang-Mills coupling.}.

The equation of motion of the metric component that follows from it takes the following form
\bea\label{metric_eom}
&&R_{MN}-\f{2\Lambda}{(d-1)}g_{MN}-\f{g_{MN}}{(d-1)}V(\phi)-\f{1}{2}\p_M\phi\p_N\phi-\nn&&\f{T_b~ Z_1(\phi)Z_2(\phi)}{4(d-1)}\f{\sqrt{-det\bigg(g~ Z_2(\phi)+\lambda F\bigg)_{ab}}}{\sqrt{-g}}\bigg[\bigg(g~ Z_2(\phi)+\lambda F\bigg)^{-1}+{\bigg(g~ Z_2(\phi)-\lambda F\bigg)}^{-1} \bigg]^{KL}\nn&&
\bigg[g_{MN}g_{KL}-(d-1)g_{MK}g_{NL} \bigg]=0.
\eea

The gauge field equation of motion is
\be\label{gauge_field_eom}
\p_M\Bigg[Z_1(\phi)\sqrt{-det\bigg(g~ Z_2(\phi)+\lambda F\bigg)_{ab}}\Bigg(\bigg(g~ Z_2(\phi)+\lambda F\bigg)^{-1}-{\bigg(g~ Z_2(\phi)-\lambda F\bigg)}^{-1}  \Bigg)^{MN}\Bigg]=0
\ee

It follows trivially that the gauge field can be fully determined in terms of the metric components and the dilaton. Finally, the equation of motion of the scalar field
\bea\label{scalar_eom}
&&\p_{M}\bigg(\sqrt{-g}\p^M\phi \bigg)-\sqrt{-g}\f{dV(\phi)}{d\phi}-T_b \f{dZ_1(\phi)}{d\phi}\sqrt{-det\bigg(g~ Z_2(\phi)+\lambda F\bigg)_{ab}}-\nn&&\f{T_b}{2}Z_1(\phi)
\sqrt{-det\bigg(g~ Z_2(\phi)+\lambda F\bigg)_{ab}}\f{dZ_2(\phi)}{d\phi}{\bigg(g~ Z_2(\phi)+\lambda F\bigg)}^{-1MN}g_{MN}=0.
\eea

Let us consider an ansatz where the metric, the abelian field strength and the dilaton to be of the following form 
\be\label{ansatz_solution}
ds^2_{d+1}=-g_{tt}(r)dt^2+g_{rr}(r)dr^2+g_{xx}(r)dx^2_i,\quad
A=A_t(r)dt,\quad F=A_t' dr\w dt,\quad \phi=\phi(r).
\ee 
In this, $r$, coordinate system, the UV is at $r\ra \infty$. The IR is at $r=r_h$ for black holes or at $r=0$  otherwise.  At IR, we define   the following form of the functions 
\be\label{def_alpha_beta}
Z_1(\phi)=exp(-\alpha\phi),\quad Z_2(\phi)=exp(\beta\phi),
\ee
and   we choose the potential as
\be\label{potential}
V(\phi)=m_1~ exp(m_2~ \phi),
\ee
where $\alpha,~\beta,~ m_1$ and $m_2$ are constants.  In which case, we shall see the dilaton goes logarithmically. Hence, diverges at IR. With this structure of the ansatz,  there exists several exact solutions.  

Given such a choice of the metric as written in eq(\ref{ansatz_solution}), the various non-vanishing components of the  Ricci tensor are
\bea
R_{tt}&=&\f{g''_{tt}}{2g_{rr}}+(d-1)\f{g'_{tt}g'_{xx}}{4 g_{rr}g_{xx}}-\f{g'^2_{tt}}{4g_{rr}g_{tt}}-\f{g'_{tt}g'_{rr}}{4g^2_{rr}},\nn
R_{ij}&=&\delta_{ij}\bigg[-\f{g''_{xx}}{2g_{rr}}-(d-3)\f{g'^2_{xx}}{4g_{rr}g_{xx}}+\f{g'_{xx}g'_{rr}}{4g^2_{rr}}-\f{g'_{tt}g'_{xx}}{4 g_{rr}g_{tt}}\bigg],\nn
R_{rr}&=&-(d-1)\f{g''_{xx}}{2g_{xx}}-\f{g''_{tt}}{2g_{tt}}+(d-1)\f{g'^2_{xx}}{4g^2_{xx}}+(d-1)\f{g'_{rr}g'_{xx}}{4 g_{rr}g_{xx}}+\f{g'^2_{tt}}{4 g^2_{tt}}+\f{g'_{tt}g'_{rr}}{4 g_{rr}g_{tt}}
\eea

Solving the equation of motion of the gauge field gives 
\be\label{sol_gauge_field}
\lambda A'_t=\f{\rho Z_2 \sqrt{g_{tt}g_{rr}}}{\sqrt{\rho^2+Z^2_1 Z^{d-1}_2g^{d-1}_{xx}}},\quad\Longrightarrow
{g_{tt}g_{rr}Z^2_2-\lambda^2 A'^2_t }=\f{g_{tt}g_{rr}g^{(d-1)}_{xx}Z^2_1 Z^{d+1}_2 }{\rho^2+Z^2_1 Z^{d-1}_2g^{d-1}_{xx}}
\ee
where  $\rho$ is the constant of integration and  interpreted as the charge density. The equation of motion of the scalar field can be simplified as 
\bea
&&\p_r\bigg(\sqrt{\f{g_{tt}}{g_{rr}}}g^{(d-1)/2}_{xx}\phi' \bigg)-\f{dV}{d\phi}\sqrt{g_{tt}g_{rr}}~ g^{(d-1)/2}_{xx}-\bigg[\f{dZ_1}{d\phi}+\f{Z_1}{Z_2}\bigg(\f{d-1}{2}\bigg) \f{dZ_2}{d\phi}\bigg]\times\nn
&&T_b  Z^{(d-1)/2}_2g^{(d-1)/2}_{xx} \sqrt{g_{tt}g_{rr}Z^2_2-\lambda^2 A'^2_t }- T_b Z_1\f{Z^{(d+1)/2}_2g_{tt}g_{rr} g^{(d-1)/2}_{xx} }{ \sqrt{g_{tt}g_{rr}Z^2_2-\lambda^2 A'^2_t }}\f{dZ_2}{d\phi}=0.
\eea

Now using the solution of the  gauge field, the equation of motion of the scalar field becomes 
\bea\label{scalar_generic_eom_dbi}
&&\p_r\Bigg(\sqrt{\f{g_{tt}}{g_{rr}}}g^{(d-1)/2}_{xx}\phi' \Bigg)- \f{dV}{d\phi}\sqrt{g_{tt}g_{rr}}~ g^{(d-1)/2}_{xx}-T_b\f{dZ_2}{d\phi}\sqrt{g_{tt}g_{rr}} \sqrt{\rho^2+Z^2_1 Z^{d-1}_2g^{d-1}_{xx}}\nn
&&-\bigg[\f{dZ_1}{d\phi}+\f{Z_1}{Z_2}\bigg(\f{d-1}{2}\bigg) \f{dZ_2}{d\phi}\bigg]\Bigg(\f{T_b\sqrt{g_{tt}g_{rr}}g^{d-1}_{xx}Z_1 Z^d_2}{\sqrt{\rho^2+Z^2_1 Z^{d-1}_2g^{d-1}_{xx}}} \Bigg)=0.
\eea

Finally, the equation of motion of the metric component can be expressed, explicitly,  as follows
\bea\label{detailed_metric_eom_dbi_gtt}
&& R_{tt}+\f{V+2\Lambda}{d-1}g_{tt}-T_b\f{(d-3)}{2(d-1)}\f{Z_2 g_{tt}}{ g^{(d-1)/2}_{xx}}{\sqrt{\rho^2+Z^2_1 Z^{d-1}_2g^{d-1}_{xx}}}+\nn&&
\quad\quad\quad\quad\quad\quad\quad\quad\quad\quad\quad\quad\quad\quad\quad\quad\quad\quad
\f{T_b}{2}\f{Z^2_1Z^{d}_2g_{tt}g^{(d-1)/2}_{xx} }{{\sqrt{\rho^2+ Z^2_1 Z^{d-1}_2g^{d-1}_{xx}}}}=0,
\eea
\bea\label{detailed_metric_eom_dbi_dilaton_gxx}
&&R_{ij}-\f{V+2\Lambda}{d-1}g_{xx}\delta_{ij}-T_b\f{\delta_{ij}}{d-1} \f{Z_2{\sqrt{\rho^2+Z^2_1 Z^{d-1}_2g^{d-1}_{xx}}}}{g^{(d-3)/2}_{xx}}=0,
%+\nn&&\quad\quad\quad\quad\quad\quad\quad\quad\quad\quad\quad\quad\quad\quad\quad\f{T_b}{2} Z^2_1 Z^d_2(1-Z_2) \f{\delta_{ij}g^{(d+1)/2}_{xx}}{\sqrt{\rho^2+Z^2_1 Z^{d-1}_2g^{d-1}_{xx}}}=0,
\eea
\bea\label{detailed_metric_eom_dbi_dilaton_grr}
&&R_{rr}-\f{V+2\Lambda}{d-1}g_{rr}-\f{1}{2}\phi'^2+T_b\f{(d-3)g_{rr}}{2(d-1)} \f{Z_2{\sqrt{\rho^2+Z^2_1 Z^{d-1}_2g^{d-1}_{xx}}}}{g^{(d-1)/2}_{xx}}-\nn&&\quad\quad\quad\quad\quad\quad\quad\quad\quad\quad\quad\quad\quad\quad\quad\quad\quad\quad\f{T_b}{2} Z^2_1 Z^{d}_2 \f{g_{rr}g^{(d-1)/2}_{xx}}{\sqrt{\rho^2+Z^2_1 Z^{d-1}_2g^{d-1}_{xx}}}=0
\eea

The equations of motion of $A_t{}(r)$ is integrable and gives rise to one independent  parameter, $\rho$. There exists $4$  unknown   functions: $g_{tt}(r),~g_{rr}(r),~g_{xx}(r)$ and $\phi(r)$, and as many equations. Hence, there exists a solution.  In the action, we have defined $3$ functions, $Z_1(\phi),~Z_2(\phi)$ and $V(\phi)$, which has   got $4$ parameters, those  are $\alpha,~\beta,~m_1,~m_2$ and there are extra $2$   parameters, $T_b$ and $\Lambda$. So all total, we have $7$ parameters and only $3$ are independent.

\section{Exact solution at IR: $AdS_2\times R^{d-1}$}

Considering a special case for which the potential energy is trivial, $V(\phi)=0$, along with constant $Z_1$ and $Z_2$, i.e., $Z_1=1=Z_2$, the dilaton can be taken as trivial. In which case, 
it is expected that the solution  near the IR end should take the following form $AdS_2\times R^{d-1}$ and  the explicit form of it  looks as
\be
ds^2_{d+1}=-\f{r^2}{R^2_2}dt^2+\f{R^2_2}{r^2}dr^2+c^2_0\delta_{ij}dx^idx^j,\quad A=\f{e_d }{R^2_2}r~dt,
\ee
where $R_2$ is the size of the $AdS_2$ spacetime and we have set $\lambda=1$, for convenience. The tension of the brane and the cosmological constant is determined as
\be\label{tension_cc_ads2}
T_b=\f{2}{e^2_d}\sqrt{R^4_2-e^2_d}, \quad 
\Lambda=-\f{T_b R^2_2}{2\sqrt{R^4_2-e^2_d}}=
-\f{R^2_2}{e^2_d}.
\ee 

It is easy to notice that for real valued  tension, $T_b$, the brane requires the constraint $R^4_2 \ge e^2_d$ and such a condition is easily met by looking at the equation of motion of the gauge field. The constant $e_d$ is determined in terms of the charge density, $\rho$, as
\be
e_d=\f{\rho ~R^2_2}{\sqrt{\rho^2+c^{2(d-1)}_0}}.
\ee
The finite temperature solution at IR with only non-zero electric field in any arbitrary $d+1$ spacetime dimensions 

\be\label{ads2Rd-1}
ds^2_{d+1}=-\f{r^2}{R^2_2}\bigg(1-\f{r_h}{r}\bigg)dt^2+\f{R^2_2}{r^2\bigg(1-\f{r_h}{r}\bigg)}dr^2+c^2_0\delta_{ij}dx^idx^j,\quad A=\f{\rho ~(r-r_h) }{\sqrt{c^{2(d-1)}_0+\rho^2}}~dt,
\ee
with the tension  of the brane and the cosmological constant as written in eq(\ref{tension_cc_ads2}).

If we want to turn on a constant magnetic field along with an electric field for which the 1-form gauge  potential takes the following form $A=\f{e_d}{R^2_2}rdt+\f{B}{2}(x_1 dx_2-x_2 dx_1)$, then 
the finite temperature solution at IR, let us say in $3+1$ spacetime dimensions,   takes the following form
\be
ds^2=-\f{r^2}{R^2_2}\bigg(1-\f{r_h}{r}\bigg)dt^2+\f{R^2_2}{r^2\bigg(1-\f{r_h}{r}\bigg)}dr^2+c^2_0(dx^2+dy^2),\quad 
A_t=\f{\rho ~(r-r_h)}{\sqrt{c^4_0+B^2+\rho^2}}.
\ee

The tension of the brane and the cosmological constant takes the following form
\be
T_b=\f{2 c^2_0\sqrt{c^4_0+B^2+\rho^2}}{R^2_2(B^2+\rho^2)},\quad \Lambda=-\f{c^4_0+B^2+\rho^2}{R^2_2(B^2+\rho^2)}.
\ee

\subsection{A black hole solution for $\beta=0,~\gamma=-1,~\delta=0$}

In $3+1$ dimensional bulk spacetime dimension, there exists a black hole solution at IR, which is conformal  to $AdS_2\times R^2$. In order to construct 
such  a black hole solution,  we choose the potential as
\be
V(\phi)=m_1~ exp(m_2~ \phi),
\ee
where $m_1$ and $m_2$ are constants. The solution reads as
\bea\label{bh_dbi_3+1_d_IR}
ds^2&=&r^2[-r^{2z}f(r)dt^2+ dx^2+dy^2+\f{dr^2}{r^2 f(r)}],\quad F=\f{\rho~ r^{z+1}}{\lambda\sqrt{1+\rho^2}}dr\w dt\nn
\phi&=&2\sqrt{z+1}~Log~r,\quad m_1=-\f{2(2+z)}{\rho^2}[z+(z+1) \rho^2],\nn
T_b&=& 2z \f{(z+2)}{\rho^2}\sqrt{1+\rho^2},\quad \alpha=\f{1}{\sqrt{z+1}},\quad\beta=0,\quad\Lambda=0\nn
\quad m_2&=&2\beta-\alpha=-\f{1}{\sqrt{z+1}},\quad f(r)=1-\bigg(\f{r_h}{r}\bigg)^{2+z},\quad Z_1=\f{1}{r^2},\quad Z_2=1,
\eea
where  $z$ is the dynamical exponent.
%Even though we have set the quantity $\Lambda$ to zero it does not mean the cosmological constant vanishes. In fact upon expanding the potential, $V(\phi)$, one sees the presence of a negative cosmological constant,  which is determined by $m_1$.
The constant $\rho$ is the charge density. In this case, the number of independent and  non-vanishing parameters are two, $\alpha$ and $\rho$.

Let us calculate the Hawking temperature associated to the black hole solution as written in eq(\ref{bh_dbi_3+1_d_IR}). It is calculated from the following formula
\be
\kappa^2=-\f{1}{2}\nabla^a \varepsilon^b\nabla_a
\varepsilon_b, \quad T_H=\f{\kappa}{2\pi},
\ee
where the null vector $\varepsilon^a$ defines the horizon, $(\varepsilon^a
\varepsilon_a)_{r_h}=0$ and the temperature, $T_H$, is evaluated on the horizon. 
For a spacetime of the form:  $ds^2_{d+1}=-g_{tt}(r) dt^2+g_{rr}(r)dr^2+g_{xx}(r)dx^idx_i$, the Hawking temperature of the system essentially  becomes
\be\label{temp_bh}
T_H=\f{1}{4\pi}\Bigg( \f{g'_{tt}}{\sqrt{g_{tt}g_{rr}}}\Bigg)_{r_h},
\ee
where prime denotes derivative with respect to the radial coordinate, $r$. Doing the calculation for the solution eq(\ref{bh_dbi_3+1_d_IR}), we find the temperature as 
\be
T_H=\f{2+z}{4\pi}r^{z}_h.
\ee
The entropy density which is the area of the horizon divided by $4G$ gives
\be
s=\f{2\pi}{\kappa^2} \Bigg(\f{4\pi}{2+z}\Bigg)^{\f{2}{z}}~ T^{\f{2}{z}}_H,
\ee
where the Newton\rq{}s constant $G$ is related to the gravitational coupling $\kappa$ as $4G=\f{\kappa^2}{2\pi}$.
It is interesting to note that the entropy density vanishes as the temperature vanishes, for finite positive  dynamical exponent. There follows, the specific heat, $c_v=T_H\bigg( \f{\p s}{\p T_H}\bigg)_{\rho}\sim T^{2/z}_H$.

Let us show the vanishing of the entropy density even in the limit of $z\ra \infty$. This essentially follows from  the argument of \cite{Hartnoll:2012wm} and is shown by considering a double scaling limit:   $z\ra \infty$ and $\theta\ra \infty$ limit with   the ratio $(-\theta/z)$ kept fixed and positive for finite spacetime dimension. For $\delta=0$, let us assume that the spacetime scales according to eq(\ref{scale_symm}). For simplicity, let us consider the following $d+1$ dimensional spacetime  
\be
ds^2_{d+1}=r^{-2\gamma}\left[-r^{2z}f(r)dt^2+dx^2_i+\f{dr2}{r^2f(r)}\right],\quad f(r)=1-(r_h/r)^{\zeta},
\ee
where $r_h$ is the horizon and $\zeta$ is a constant. In this case the Hawking temperature goes as $T_H\sim r^{z}_h$, whereas the entropy goes as $S\sim r^{-\gamma(d-1)}_h\sim T^{-(d-1)\gamma/z}_H=T^{-\theta/z}_H$. From this expression of the entropy, it follows that the entropy  vanishes in the  $z\ra \infty$ and $\theta\ra \infty$ limit with   the ratio $(-\theta/z)$ kept  fixed and positive  at low temperature.

For finite $\theta$ and $z\ra\infty$, the vanishing of the entropy density at low temperature is  subtle. 
%Essentially, it depends on the order of the limit.
In what follows, for finite $\theta$, we shall be taking $z$ finite as well.
%So, we shall consider taking first $T_H\ra 0$ and then $z\ra\infty$.

\subsubsection{Null Energy Condition}

Given the choice of our action in eq(\ref{eh_dbi_dilaton}), the energy-momentum tensor takes the following form
\bea
T_{MN}&=&\p_M\phi\p_N\phi-g_{MN}\Bigg[V+2\Lambda+\f{1}{2}(\p\phi)^2\Bigg]-\nn&&\f{T_b}{2\sqrt{-g}} Z_1 Z_2\sqrt{-det(Z_2 g+\lambda F)} 
\bigg[(Z_2 g+\lambda F)+(Z_2 g-\lambda F)\bigg]^{KL}g_{KM}g_{NL}
\eea

Demanding that the system  we are dealing with should satisfy the null energy condition, $T_{MN}u^Mu^N \geq 0$ for some null vectors $u^M$ gives us the restriction on the Ricci tensor as $R_{MN}u^Mu^N \geq 0$.   By considering the two possible choices for the null vectors
as $u^t=1/\sqrt{g_{tt}},~u^r=1/\sqrt{g_{rr}},u^i=0$ and   $u^t=1/\sqrt{g_{tt}},~u^{x_1x_1}=1/\sqrt{g_{x_1x_1}}, ~u^r=0$
 and setting the rest of the vectors to  zero, gives the following conditions for the metric of the type 
\be
ds^2_{d+1}=-r^{2(z+1)} dt^2+r^2 dx^idx_i+dr^2,\quad \Rightarrow (z+1)(d-1)\geq 0,\quad z(d+z-1)\geq 0.
\ee
We obtain such a  form of the metric  in the  zero temperature limit of eq(\ref{bh_dbi_3+1_d_IR}). 
Upon solving the inequality, we find the most interesting restriction that  is
\be
d\geq 2\quad z\geq 0,
\ee
whereas the other possibilities are not that interesting because either  the dimensionality of the spacetime or  the dynamical exponent could become negative.

\subsection{Lifshitz solution: $(\alpha\neq 0,~ \beta\neq 0,~\gamma=0,~\delta=1)$}

In order to generate a Lifshitz solution, we shall consider the case where all the degrees of freedom are non-trivial i.e., they do not vanish, as well as the functions $Z_1$ and $Z_2$ are not set to unity. But we shall take a trivial potential energy, $V=0$ with non-zero cosmological constant, $\Lambda\neq 0$. For simplicity, we shall be solving the equations of motion in $3+1$ dimensional bulk spacetime dimensions. In this case, the solution reads as
\bea\label{lifshitz_sol}
ds^2_{3+1}&=&-r^{2z}f(r) dt^2+r^2(dx^2_1+dx^2_2)+\f{dr^2}{r^2f(r)},\quad f(r)=1-\f{r^{z+2}_h}{r^{z+2}},\nn
\phi(r)&=&2\sqrt{z-1}~Log~r,\quad Z_1=\f{1}{r^4},\quad Z_2=r^2,\quad F=\f{\rho r^{z+1}}{\sqrt{1+\lambda^2\rho^2}}dr\w dt\nn
T_b&=&2\f{\sqrt{1+\lambda^2\rho^2}}{\lambda^2\rho^2}(z^2+z-2),\quad \Lambda=-\f{z^2(1+\lambda^2\rho^2)+z(1+2\lambda^2\rho^2)-2}{\lambda^2\rho^2},\nn
\alpha&=&\f{2}{\sqrt{z-1}},\quad \beta=\f{1}{\sqrt{z-1}},\quad \gamma=0,
\eea 
where  the dynamical exponent  $z$  should always be bigger than unity, $z > 1$.  In this case, the number  non-vanishing and  independent  parameters are two, $\alpha$ and $\rho$.
The Hawking temperature, in this case, turns out to be
\be
T_H=\f{(z+2)}{4\pi}r^z_h.
\ee
With the Bekenstein-Hawking entropy density  given as $s=\kappa^2/(2\pi)~ (\f{4\pi}{(z+2)})^{2/z} T^{2/z}_H$.  From this expression of the entropy density, it follows trivially that for finite and  positive dynamical exponent the entropy vanishes as temperature vanishes. The specific heat, $c_v=T_H\bigg( \f{\p s}{\p T_H}\bigg)_{\rho}\sim T^{2/z}_H$. It is interesting to note that for a specific choice of the dynamical exponent, $z=2$, the specific heat has a linear temperature dependence.

\section{Solutions at IR in arbitrary spacetime dimensions}

In this section, we shall generalize the solutions found in the previous section to arbitrary spacetime dimensions.
Instead of giving the details, let us write down the solution in $d+1$ dimensional spacetime,  at IR, to the eq(\ref{scalar_generic_eom_dbi}) and eq(\ref{detailed_metric_eom_dbi_gtt})-eq(\ref{detailed_metric_eom_dbi_dilaton_grr}) with the scaling symmetry as written down in 
eq(\ref{scale_symm}).

\underline{ Conformal to $AdS_2\otimes R^{d-1}$, i.e.,  for $\delta=0,~\gamma\neq 0$:}\\
 
The solution reads as
\bea
ds^2_{d+1,\delta=0}&=&r^{-2\gamma}\left[-r^{2z}f(r)dt^2+dx^idx^j\delta_{ij}+\f{dr^2}{r^2f(r)}\right],\quad f(r)=1-\left(r_h/r\right)^{\eta},\nn
\phi(r)&=&\sqrt{2\gamma(d-1)(\gamma-z)}~Log~r,\quad \eta=z-\gamma(d-1),\quad \Lambda=0,\nn
\alpha&=&(d-1)^{3/2} \sqrt{\f{\gamma}{8(\gamma-z)}},\quad \beta=(d-3)\sqrt{\f{\gamma}{2(d-1)(\gamma-z)}},
\nn
 T_b&=&\f{2z}{\rho^2}[z-\gamma(d-1)]\sqrt{1+\rho^2},\quad m_1=-\f{(z-\gamma(d-1))}{\rho^2}[2z(1+\rho^2)-\gamma\rho^2(d-1)],\nn
 V(r)&=&-\f{(z-\gamma(d-1))}{\rho^2}[2z+(2z+\gamma(1-d))\rho^2]r^{2\gamma},\quad m_2=\sqrt{\f{2\gamma}{(d-1)(\gamma-z)}},\nn
 A\rq{}_t&=&\f{\rho}{\lambda\sqrt{1+\rho^2}}r^{z-1-\gamma(d-1)}
\eea

For this solution the Hawking temperature and the Bekenstein-Hawking entropy density becomes
\be\label{entropy_delta_0}
T_H=\f{\eta}{4\pi}r^z_h,\quad s\sim \f{2\pi}{\kappa^2}~ T^{\f{-\gamma(d-1)}{z}}_H.
\ee

The boundary is at $r=\infty$, which means $\gamma<0$ and $z-\gamma>0$. These two conditions suggests that $\gamma <0$ and 
 $z\geq 0$. The null energy condition (NEC) gives the constraint as $\gamma(\gamma-z)\geq 0$ and $z(z-\gamma(d-1))\geq 0$. It means for $d> 1$, the NEC suggests to consider $z\geq 0$ also.
 
 To  get the solution for  $AdS_2\otimes R^{d-1}$, we need to set $\gamma=0$, in  which case there exists  a non-zero entropy density and  studied in detail e.g., in \cite{Faulkner:2009wj}.\\

\underline{ Conformal to Lifshitz solution, i.e.,  for $\delta=1,~\gamma\neq 0$:}\\
 
The solution reads as
\bea
ds^2_{d+1,\delta=1}&=&r^{-2\gamma}\left[-r^{2z}f(r)dt^2+r^2 dx^idx^j\delta_{ij}+\f{dr^2}{r^2f(r)}\right],\quad f(r)=1-\left(r_h/r\right)^{\eta},\nn
\phi(r)&=&\sqrt{2(1-\gamma)(d-1)(z-\gamma-1)}~Log~r,\quad \eta=z+(1-\gamma)(d-1),\quad \Lambda=0,\nn
\alpha&=&[d+1+\gamma(1-d)]\sqrt{\f{(d-1)}{8(1-\gamma)(z-\gamma-1)}}
,\quad \beta=\f{(d-1+\gamma(3-d))}{\sqrt{2(d-1)(1-\gamma)(z-\gamma-1)}},\nn
 T_b&=&\f{2(z-1)}{\rho^2}[z-\gamma(d-1)+d-1]\sqrt{1+\rho^2},\quad m_2=\f{\sqrt{2}\gamma}{\sqrt{(d-1)(1-\gamma)(z-\gamma-1)}},\nn
 m_1&=& -\f{1}{\rho^2}[z+(1-\gamma)(d-1)][2z(1+\rho^2)-2+\rho^2(\gamma(1-d)+d-3)],\nn
 V(r)&=&-\f{(z+(1-\gamma)(d-1))}{\rho^2}[2z(1+\rho^2)-2+(d-3+\gamma(1-d))\rho^2]r^{2\gamma},\nn
 A\rq{}_t&=&\f{\rho}{\lambda\sqrt{1+\rho^2}}r^{z+d-2-\gamma(d-1)}
\eea

In order to have the boundary at $r=\infty$, we must impose the following conditions $z-\gamma>0$ and $1-\gamma >0$. For $d>1$, the reality of the solution  implies $z-\gamma>1$. The NEC of the zero temperature limit of the solution imposes the following conditions $(d-1)(1-\gamma)(z-1-\gamma)\geq 0$ and $(z-1)(z+(d-1)(1-\gamma))\geq 0$. Combining all these constraints we have for $d>1$ as $1-\gamma >0,~z-\gamma>1$ and $z>1$.

Now the  Hawking temperature and the Bekenstein-Hawking entropy density becomes
\be\label{entropy_delta_1}
T_H=\f{\eta}{4\pi}r^z_h,\quad s\sim \f{2\pi}{\kappa^2}~ T^{\f{(1-\gamma)(d-1)}{z}}_H.
\ee

\underline{  Lifshitz solution, i.e.,  for $\delta=1,~\gamma= 0$:}\\
 
The solution reads  for trivial potential,  $V(\phi)=0$, as
\bea\label{lifshitz_sol_generic}
ds^2_{d+1}&=&-r^{2(z+1)}f(r) dt^2+r^2dx^2_i+\f{dr^2}{r^2f(r)},\quad f(r)=1-\left(\f{r_h}{r}\right)^{z+d},\nn
\phi(r)&=&\sqrt{2z(d-1)}~Log~r,\quad Z_1=r^{-\f{(d^2-1)}{2}},\quad Z_2=r^{d-1},\nn 
F&=&\f{\rho r^{d+z-1}}{\sqrt{1+\lambda^2\rho^2}}dr\w dt,\quad 
T_b=2z(d+z)\f{\sqrt{1+\rho^2}}{\rho^2},\nn
 \Lambda&=&-\left[\f{(d^2+dz-z-1)\rho^2+2z(d+z)(1+\rho^2)}{2\rho^2}\right],
\eea 
where  the dynamical exponent  $z$  should always be bigger than zero, $z \geq  0$ for $d>1$, which follows from the NEC. The entropy density of the solution goes as $s \sim T^{\f{(d-1)}{z}}_H$, where $T_H$ is the Hawking temperature.\\

\underline{ General expression of the entropy}\\

The entropy densities  in these  cases,  eq(\ref{entropy_delta_0}),  eq(\ref{entropy_delta_1}) and eq(\ref{lifshitz_sol}) can be written together for any choice of $\delta$ and $\gamma$ as
\be\label{gen_exp_entropy}
s\sim \f{2\pi}{\kappa^2}~ T^{\f{(\delta-\gamma)(d-1)}{z}}_H.
 \ee 

Let us redefine, $\gamma=\f{\theta}{(d-1)}$, in which case the above mentioned entropy density can be re-written as
 \be
 s\sim \f{2\pi}{\kappa^2}~ T^{\f{\delta(d-1)-\theta}{z}}_H,
 \ee
which is the expression of the entropy density as suggested in eq(\ref{generic_entropy}). In fact this form of the entropy density generalizes  the one as suggested for $\delta=1$ case in \cite{Huijse:2011ef}.

Let us recall that in scale invariant model with dynamical exponent $z$, the entropy density goes as $s\sim \f{2\pi}{\kappa^2}~ T^{\f{(d-1)}{z}}_H$. Upon comparing with eq(\ref{gen_exp_entropy}), it follows that  as far as the entropy density is concerned
there  won\rq{}t be any distinction between the scale symmetry violating or preserving theories for 
\be
\delta=1+\gamma.
\ee
This equation is satisfied for two choices: $(1)~\gamma=0,~\delta=1$ and   $(2)~\gamma=-1,~\delta=0$, as $\delta$ can take only two values. The first type belongs to the Lifshitz solution whereas the second type belongs to the solution that are conformal to $AdS_2\otimes R^{d-1}$.

\section{Black hole solution at UV}

Let us construct a black hole solution for a specific choice of the functions that appear in the action eq(\ref{eh_dbi_dilaton}), namely,  we set $Z_1=1=Z_2$ and the potential energy as $V=0$. 
For this choice of the functions, it follows  that  the solution of the scalar field can be taken as trivial i.e., $\phi=0$, in which case the  
 gauge field takes the following form
\be\label{gauge_field_dbi}
\lambda A'_t=\f{\rho\sqrt{g_{tt}g_{rr}}}{\sqrt{\rho^2+g^{d-1}_{xx}}}.
\ee 

The equation of motion of the metric component reduces to 
\bea\label{metric_eom_dbi}
&&R_{MN}-\f{2\Lambda}{(d-1)}g_{MN}-\f{T_b}{4(d-1)}\f{\sqrt{-det\bigg(g+\lambda F\bigg)_{ab}}}{\sqrt{-g}}\bigg[\bigg(g+\lambda F\bigg)^{-1}+{\bigg(g-\lambda F\bigg)}^{-1} \bigg]^{KL}\nn&&
\bigg[g_{MN}g_{KL}-(d-1)g_{MK}g_{NL} \bigg]=0.
\eea

Let us assume that the geometry,  asymptotically, approach   the $AdS$ spacetime.   We consider   the following form of the spacetime for explicit calculations
\be\label{sol_UV_electrically_d+1}
ds^2_{d+1}=\f{r^2}{R^2}[-f(r)dt^2+dx^2_i]+\f{R^2 dr^2}{r^2 f(r)}, 
\ee
where $R$ is the size of the $AdS$ spacetime. Let us substitute this ansatz into the equations of motion of the metric eq(\ref{metric_eom_dbi}), then there arises two second order   differential equations. One from the $g_{tt}$ and the other  from the $g_{rr}$ component. In fact these two differential equations are not independent, the precise relation is $\f{2R^4}{r^2f}\times$ eq(\ref{detailed_metric_eom_dbi_gtt}) $=-2r^2\times$ eq(\ref{detailed_metric_eom_dbi_dilaton_grr}).
 So we left with only one second order differential equation, which reads as  
\bea\label{eom_f}
r^2f''+(d+3)r f'+2df+\f{4\Lambda R^2}{d-1}+\f{ T_b R^2 (r/R)^{d-1}}{\sqrt{\rho^2+ \f{r^{2(d-1)}}{R^{2(d-1)}}}}-\f{d-3}{d-1}T_b R^2  \f{r^{d-1}}{R^{d-1}} \sqrt{\rho^2+ \f{r^{2(d-1)}}{R^{2(d-1)}}}=0.
\eea

Now, this equation can be reduced to a first order differential equation, which essentially follows from eq(\ref{detailed_metric_eom_dbi_dilaton_gxx}) and the precise relation is $-\f{R^4}{r}\p_r (eq(\ref{detailed_metric_eom_dbi_dilaton_gxx}))=$ eq(\ref{eom_f}).  Finally, the equation of motion that follows from eq(\ref{detailed_metric_eom_dbi_dilaton_gxx})
\be\label{1st_order_eom}
rf'(r)+d f(r)+\f{2\Lambda R^2}{d-1}+\f{T_b R^2}{d-1}\f{r^{1-d}}{R^{1-d}} \sqrt{\rho^2+ \f{r^{2(d-1)}}{R^{2(d-1)}}}=0.
\ee
On solving this differential equation for the generic choice of the dimension gives 
\be\label{sol_f_without_scalar}
f(r)=\f{c_1}{r^d}-\f{2\Lambda R^2}{d(d-1)}-\f{T_b R^2\rho}{(d-1)}\f{r^{1-d}}{R^{1-d}}~~~{}_2F_1\bigg[-\f{1}{2},\f{1}{2(d-1)},\f{2d-1}{2(d-1)},-\f{r^{2(d-1)}}{R^{2(d-1)}\rho^2}\bigg].
\ee

To get a feel of the solution,   in what follows, we shall try to solve it for few specific choices of the spacetime dimension. Let us assume that the geometry asymptotes to $AdS_3$, i.e., we set $d=2$, in which case  the solution is
\bea
ds^2_3&=&\f{r^2}{R^2}[-f(r)dt^2+dx^2]+\f{R^2 dr^2}{r^2 f(r)},  \quad {\rm with}\nn
f(r)&=&\f{c_1}{r^2}-\Lambda R^2-T_b R^2\f{\sqrt{r^2+R^2\rho^2}}{2r}-\f{R^4 T_b\rho^2}{2r^2}Log\bigg(r+\sqrt{r^2+R^2\rho^2}\bigg),
\eea
where $c_1$ is a constant. The horizon, $r_h$, is determined as the location for which $f(r_h)=0$. For $AdS_4$, the solution looks as
\bea\label{ads4_sol_e_dbi}
ds^2_4&=&\f{r^2}{R^2}[-f(r)dt^2+dx^2_1+dx^2_2]+\f{R^2 dr^2}{r^2 f(r)}, \nn 
f(r)&=&\f{c_1}{r^3}-\f{1}{3}R^2\Lambda-\f{1}{6r^2} T_bR^2\sqrt{r^4+R^4\rho^2}-\f{1}{3r^2}T_b\rho R^4~ {}_2F_1[\f{1}{4},\f{1}{2},\f{5}{4},-\f{r^4}{R^4\rho^2}].
\eea 

Similarly, solving for an $AdS_5$ spacetime, we find  
\bea
ds^2_5&=&\f{r^2}{R^2}[-f(r)dt^2+dx^2_1+dx^2_2+dx^2_3]+\f{R^2 dr^2}{r^2 f(r)}, \nn 
f(r)&=&\f{c_1}{r^4}-\f{1}{6}R^2\Lambda-\f{1}{12r^3} T_bR^2\sqrt{r^6+R^6\rho^2}-\f{1}{4r^3}T_b\rho R^5~ {}_2F_1[\f{1}{6},\f{1}{2},\f{7}{6},-\f{r^6}{R^6\rho^2}],
\eea 
where ${}_2F_1[a,b,c,x]$ is the hypergeometric function. In order to fix the  constant, $c_1$,  we need to do an expansion in the small charge density, $\rho$, limit and compare it with the RN black hole solution. From which we can identify the constant $c_1$  as the mass density of the black hole,  $c_1\propto -M$. The explicit identification is presented towards the end of this section.

The gauge field that supports the  $AdS$ spacetime, from eq(\ref{gauge_field_dbi}), follows as
\be
 A_t(r)=\f{r}{\lambda}~~{}_2F_1\Bigg[\f{1}{2},\f{1}{2(d-1)},\f{2d-1}{2(d-1)},-\f{(r/R)^{2(d-1)}}{\rho^2}\Bigg]+\f{2\kappa^2}{\lambda T_b}{\rm \Phi}.
\ee 
The constant quantity, $\Phi$,  is determined by requiring that the gauge field should vanish at the horizon, $A_t(r_h)=0$, in order to keep the norm of the gauge potential  finite at the horizon.  
The chemical potential is determined by 
\bea
&&\mu=\int^{\infty}_{r_h}~ A'_t=\f{1}{\lambda}\int^{\infty}_{r_h} \f{\rho R^{d-1}}{\sqrt{r^{2(d-1)}+\rho^2 R^{2(d-1)}}}\nn
&&=\f{1}{\lambda}\left(\f{\rho_t^{\f{1}{d-1}}}{(d-2)\sqrt{\pi}}\Gamma\bigg(\f{4-3d}{2-2d}\bigg)\Gamma\bigg(\f{1}{2d-2}\bigg)-r_h ~{}_2F_1\Bigg[\f{1}{2},\f{1}{2(d-1)},\f{1-2d}{2-2d},-\f{r^{2(d-1)}_h}{\rho^2_t}\Bigg] \right),\nn
\eea
where $\rho_t\equiv \rho R^{d-1}$. This particular form of the gauge potential, hence the chemical potential, matches precisely with the one computed in the probe approximation in \cite{Karch:2008fa} i.e., without taking the back reaction of the gauge field onto the geometry. We can determine  whether this particular state corresponds to a compressible phase or not by simple looking at the continuity of the chemical potential with respect to the charge density. 
\bea
\f{d\mu}{d\rho}&=&\f{R}{\lambda(d-1)^2}\Bigg(\f{\rho^{\f{2-d}{d-1}}}{\sqrt{\pi}}\Gamma\bigg(\f{4-3d}{2-2d}\bigg)\Gamma\bigg(\f{1}{2d-2}\bigg)+\f{r_h\rho (d-1)}{\sqrt{\rho^2-R^{2-2d}r^{2d-2}_h}}\nn&&-r_h(d-1) ~{}_2F_1\Bigg[\f{1}{2},\f{1}{2(d-1)},\f{1-2d}{2-2d},\f{r^{2(d-1)}_hR^{2(1-d)}}{\rho^2}\Bigg] \Bigg).
\eea

At a very specific value of the charge density, namely, $\rho=(r_h/R)^{d-1}$, the above derivative has a singularity.  So, we conclude that the dual field theory of the  Einstein-DBI system does not show up the necessary feature to be part of the compressible phase of matter. 

The temperature of the $d+1$ dimensional  black hole can be computed from the formula as written in eq(\ref{temp_bh}) as
\be
T_H=-\f{ r_h}{(d-1)4\pi }\Bigg[2\Lambda+T_b~ r^{1-d}_h\sqrt{\rho^2R^{2(d-1)}+r^{2(d-1)}_h}\Bigg],
\ee
where we have used eq(\ref{1st_order_eom}) to find the derivative of the function $f(r)$. The Bekenstein- Hawking entropy density becomes
\be
s=\f{2\pi}{\kappa^2}\Bigg( \f{r_h}{R}\Bigg)^{d-1}.
\ee

Let us find the entropy in a limit for which the temperature of the black hole vanishes, $T_H=0$,  for a non-zero size of the horizon, $r_h\neq 0$. This happen when the size of the horizon takes the following form
\be
r^{d-1}_h=\f{T_b \rho R^{d-1}}{ \sqrt{4\Lambda^2-T^2_b}}.
\ee

In this case the entropy density becomes
\be
s_{ext}=\f{2\pi}{\kappa^2}\f{T_b \rho }{ \sqrt{4\Lambda^2-T^2_b}}  \quad \neq 0.
\ee

It means even for the non-linearly generalized  Einstein-Maxwell action that is the Einstein-DBI action has a non-zero entropy at zero temperature.  Moreover, the existence of non-zero entropy or the non-zero horizon size at zero temperature suggests   an upper  bound on the tension of the brane $T^2_b < 4\Lambda^2$ and is consistent with the solution  found in eq(\ref{tension_cc_ads2}) for $AdS_2$ .

The specific heat $C_v=T_H\bigg(\f{\p s}{\p T_H}\bigg)_{\rho}$ that follows
\be
C_v=\f{2(d-1)\pi R^{3-d}r^d_h  \sqrt{\rho^2 R^{2(d-1)}+r^{2(d-1)}_h}[2\Lambda r^d_h+T_b r_h  \sqrt{\rho^2 R^{2(d-1)}+r^{2(d-1)}_h}]}{\kappa^2[R^2 r^d_h(T_b r^d_h+2\Lambda r_h \sqrt{\rho^2 R^{2(d-1)}+r^{2(d-1)}_h})-(d-2) T_b \rho^2 R^{2d}r^2_h]}
\ee

In our notation $T_b$ is positive and we are dealing with spacetimes  of negative cosmological constant, which  means there exists a range of values of the  charge density for which the solution has got positive specific heat.  In this range of charge densities,  the system looks to be  thermodynamically stable. In fact,
 for  a choice like, $d=3, ~R=1=r_h$, the specific heat, $C_v=\f{4\pi\sqrt{1+\rho^2}[2\Lambda+T_b \sqrt{1+\rho^2}] }{\kappa^2[T_b(1-\rho^2)+2\Lambda\sqrt{1+\rho^2}]}$. For small charge density, it becomes $C_v=\f{4\pi}{\kappa^2}+\f{8\pi T_b}{\kappa^2[T_b+2\Lambda]}\rho^2+{\cal O}(\rho)^4.$   

Let us fix the precise relation between the constant $c_1$ and the mass  density $M$, in order to do so,  let us use the thermodynamic relation $dM=T_H ds+\Phi d\rho$. Since the charge  density is constant means the mass of the black hole can be found from 
$M=\int dr_h T_H \bigg(\f{\p S}{\p r_h}\bigg)$. Doing the above integral along with the use of the following  relations  for Hypergeometric functions 
\be
(a-b)~ {}_2F_1[a,b,c,x]=a~{}_2F_1[1+a,b,c,x]-b~{}_2F_1[a,1+b,c,x],\quad {}_2F_1[a,b,b,x]=(1-x)^{-a},
\ee
gives the mass as
\be
-2\kappa^2 R^{d-1} M=\f{2\Lambda}{d}  r^d_h+T_b \rho R^{d-1}r_h ~{}_2F_1\Bigg[-\f{1}{2},\f{1}{2(d-1)},\f{1-2d}{2-2d},-\f{r^{2(d-1)}_h}{\rho^2R^{2(d-1)}}\Bigg].
\ee

 Recall that the constant $c_1$ is determined from the condition, $f(r_h)=0$, which  means
\be
c_1=\f{2\Lambda R^2 r^d_h}{d(d-1)}+\f{T_b \rho R^{d+1}}{(d-1)}r_h ~{}_2F_1\Bigg[-\f{1}{2},\f{1}{2(d-1)},\f{1-2d}{2-2d},-\f{r^{2(d-1)}_h}{\rho^2R^{2(d-1)}}\Bigg].
\ee

On comparing these two expressions, we find $c_1=-\f{2\kappa^2}{(d-1)} R^{d+1}M$.

\subsubsection{Stability}

The thermodynamic stability condition as suggested by the Gubser-Mitra conjecture \cite{Gubser:2000mm} requires that 
\be\label{stability_Hes}
 det~\left(\f{\p^2M}{\p(s,\rho)^2} \right)=\f{\p^2 M}{\p s^2} \f{\p^2 M}{\p \rho^2} -\left(\f{\p M}{\p s} \f{\p M}{\p \rho} \right)^2 > 0.
\ee

Let us define few dimensionless objects  with subscript $0$ : $T_{b0}=T_b~\kappa^{4/(d-1)},~s_0=s~\kappa^2$ and $\Lambda_0=\Lambda \kappa^{4/(d-1)}$ and then upon 
 computing eq(\ref{stability_Hes}) gives for small $s_0/\rho\equiv X\ll 1$
 \be
 2^{\f{(2-4d)}{(d-1)}}\pi^{-2\f{(d+1)}{(d-1)}}\f{R^2 T^2_{b0}}{(d-1)^2(2d-1)}(X\rho)^{\f{(4-2d)}{(d-1)}}\left[(d+1) \pi^{\f{2}{d-1}}X^2-(8d-4)\pi^{\f{2d}{d-1}} +\cdots\right].
 \ee
 
 It just follows that the coefficient of the leading term is negative, which suggests that it is unstable. Let us look at the sign of the specific heat as argued in  \cite{Gubser:2000mm}.  The positive specific heat requires:  $\f{1}{(\p M/\p s)^3}~\f{\p^2 M}{\p s^2} >0$. Upon doing the calculation using the above mentioned dimensionless variables in $3+1$ dimensional spacetime for small $X$, means high charge density limit
 \be
 \f{1}{(\p M/\p s)^3}~\f{\p^2 M}{\p s^2}\simeq -\f{\kappa^6}{R^2\rho^2 T^2_{b0}}\bigg[16\pi-64 \f{\Lambda_0}{T_{b0}}X+{\cal O}(X)^2\bigg].
 \ee

In the low charge density limit, $\rho/s_0\equiv Y \ll 1$, the quantity
\be
\f{1}{(\p M/\p s)^3}~\f{\p^2 M}{\p s^2}\simeq \f{64\pi^3\kappa^6}{R^2s^2_0(T_{b0}+2\Lambda_0)^2}+{\cal O}(Y)^2
\ee

For asymptotically AdS spacetime, it  means the specific heat is positive for small charge density, which makes the analysis consistent as done in the previous section. So to conclude  the specific heat is positive and the quantity $ det~\left(\f{\p^2M}{\p(s,\rho)^2} \right)$  is positive for $ T_{b0}+2\Lambda_0 <0$  in the small charge density limit. Whereas in the limit of the ratio of  large charge density to entropy density  makes $\f{1}{(\p M/\p s)^3}~\f{\p^2 M}{\p s^2}$   negative, which is again consistent with the fact that in the large charge density limit the specific heat becomes negative, implying instability.

\subsection{Dyonic solution}

In this case we turn on both the electric field along with a magnetic field in $3+1$ spacetime dimensions. The magnetic field is considered to be  constant, for simplicity. Hence, the explicit structure of the field strength and the metric  is  
\be\label{dyonic_field_strength}
F=A'_t(r) dr\w dt+B dx\w dy,\quad ds^2_{3+1}=-g_{tt}(r) dt^2+g_{xx}(r)(dx^2+dy^2)+g_{rr}(r)dr^2.
\ee

Let us solve the equation of motion associated to the gauge field  and is given as 
\be\label{dyonic_gauge_field_sol}
\lambda A'_t=\f{\rho Z_2 \sqrt{g_{tt}g_{rr}}}{\sqrt{\rho^2+Z^2_1 (Z^2_2g^2_{xx}+\lambda^2 B^2)}},\quad\Longrightarrow
{g_{tt}g_{rr}Z^2_2-\lambda^2 A'^2_t }=\f{g_{tt}g_{rr}(Z^2_2g^2_{xx}+\lambda^2 B^2)Z^2_1 Z^2_2 }{\rho^2+Z^2_1 (Z^2_2g^2_{xx}+\lambda^2 B^2)}.
\ee

With a non-trivial magnetic and electric  field the equation of motion of the metric components gets modified and  are given as
\bea
 && R_{tt}+\f{V+2\Lambda}{2}g_{tt}+
\f{T_b}{2}\f{Z^2_1Z^{2}_2g_{tt}g_{xx} }{{\sqrt{\rho^2+Z^2_1 (Z^2_2g^2_{xx}+\lambda^2 B^2)}}}=0,
\eea
\bea\label{detailed_metric_eom_dbi_gxx_b}
&&R_{ij}-\f{V+2\Lambda}{2}g_{xx}\delta_{ij}-T_b\f{\delta_{ij}}{2} Z_2{\sqrt{\rho^2+Z^2_1 (Z^2_2g^2_{xx}+\lambda^2 B^2)}}=0,
\eea
\bea\label{detailed_metric_eom_dbi_grr_b}
&&R_{rr}-\f{V+2\Lambda}{2}g_{rr}-\f{1}{2}\phi'^2-\f{T_b}{2} Z^2_1 Z^{3}_2 \f{g_{rr}g_{xx}}{\sqrt{\rho^2+Z^2_1 (Z^2_2g^2_{xx}+\lambda^2 B^2)}}=0.
\eea

Now, we shall consider a specific configuration for which  the potential energy is taken as trivial, $V=0$ and $Z_1=1=Z_2$. In this case, again, the trivial dilaton profile is a solution, $\phi=0$, to the equation of motion. In order to find the black hole solution, let us demand that the solution asymptotically looks as an AdS spacetime. In which case, the following ansatz to the metric solves the equations of motion
\be\label{dyonic_dbi}
ds^2_{3+1}=\f{r^2}{R^2}\bigg[-f(r) dt^2+dx^2+dy^2\bigg]+\f{R^2}{r^2}\f{dr^2}{f(r)},
\ee
with some form for the function $f(r)$. From the $g_{tt}$ and the  $g_{rr}$ part of the metric components we find the following second order differential equation for the function $f(r)$
\be
r^2f''(r)+6rf(r)+6f(r)+2R^2\Lambda+ \f{T_b r^2 R^2}{\sqrt{r^4+R^4(\rho^2+\lambda^2 B^2)}}=0,
\ee
whereas from the $g_{xx}$ component of the metric, we find the following first order differential equation for $f(r)$
\be\label{diff_dyonic_dbi}
rf'(r)+3f(r)+\Lambda R^2+\f{T_b R^2}{2r^2}\sqrt{r^4+R^4(\rho^2+\lambda^2 B^2)}=0.
\ee
One can easily check that these two differential equations are not independent of each other. On solving the first order differential equation, we find the function, $f(r)$, has the following form  
\be
f(r)=\f{c_1}{r^3}-\f{\Lambda R^2}{3}-T_b R^2\f{\sqrt{r^4+R^4(\rho^2+\lambda^2 B^2)}}{6r^2}-\f{T_b R^4\sqrt{\rho^2+\lambda^2 B^2}}{3r^2} {}_{2}F_{1}\bigg[\f{1}{4},\f{1}{2},\f{5}{4},-\f{r^4}{R^4(\rho^2+\lambda^2 B^2)}\bigg].
\ee

It is easy to see that this solution in the zero magnetic field limit reproduces eq(\ref{ads4_sol_e_dbi}). As was done for the solution   eq(\ref{ads4_sol_e_dbi}), we can identify the constant $c_1\propto -M$, as the mass density of the black hole.  The horizon is determined from the zero's of the function, $f(r_h)=0$ and the entropy density is given as $s=(2\pi/\kappa^2)(r_h/R)^2$ . 

The temperature of such a dyonic solution is
\be
T_H=-\f{ r_h}{8\pi }\Bigg[2\Lambda+\f{T_b}{ r^2_h}\sqrt{(\rho^2+\lambda^2 B^2)R^{4}+r^{4}_h}\Bigg],
\ee 
 and the entropy in the extremal limit takes the following form
\be
s_{ext}=\f{2\pi}{\kappa^2}\f{T_b \sqrt{\rho^2+\lambda^2 B^2} }{ \sqrt{4\Lambda^2-T^2_b}}  \quad \neq 0.
\ee

There arises an interesting question: Is it possible that $4\Lambda^2=T^2_b$ ? In which case the entropy in the extremal limit diverges. In order to answer such a question, let us look at the relation between the cosmological constant and the tension of the brane.  From the condition of the vanishing temperature, there follows 
\be
\f{4\Lambda^2}{T^2_b}=1+\f{R^{4}(\rho^2+\lambda^2 B^2)}{r^{4}_h}\quad \Longrightarrow\quad  \f{4\Lambda^2}{T^2_b}\quad > \quad  1.
\ee  

So, in the vanishing temperature limit  the magnitude of the cosmological constant should be  bigger than half the tension of the brane. 

The chemical potential for such a dyonic solution is determined as
\bea
&&\mu=\int^{\infty}_{r_h}~dr~  A'_t=\f{1}{\lambda}\int^{\infty}_{r_h}dr \f{\rho R^2}{\sqrt{r^{4}+(\rho^2+\lambda^2 B^2) R^{4}}}\nn
&&=\f{\rho R^2}{\lambda}\Bigg(\f{4}{R(\rho^2+\lambda^2 B^2)^{1/4}\sqrt{\pi}}\Gamma^2\bigg(\f{5}{4}\bigg)-r_h ~{}_2F_1\Bigg[\f{1}{2},\f{1}{4},\f{5}{4},-\f{r^{4}_h}{R^4(\rho^2+\lambda^2 B^2)}\Bigg] \Bigg),\nn
\eea

\subsection{Connection with the BI black hole}

Recently, an electrically charged black hole solution is found in arbitrary spacetime dimension with the Born-Infeld (BI) matter \cite{Dey:2004yt}, see also \cite{Fernando:2003tz}-
\cite{Myung:2012np}.  It is certainly interesting to find the connection between the Einstein-DBI black hole solutions for trivial dilaton with  that of the Einstein-BI black holes in the presence of a cosmological constant. The BI matter and the DBI matter is described  
\be
S_{BI}=\int \sqrt{-g}\sqrt{1+\alpha F^{MN}F_{MN}},\quad S_{DBI}=\int \sqrt{-det(g+\lambda F)_{MN}}
\ee
where $\alpha$ is a parameter.  For small $\alpha$  we  see  the Maxwellian structure. 

Generically, the DBI action is completely different from the BI action\footnote{Upon expanding the determinant in the DBI matter for arbitrary spacetime dimension $det(g+\lambda F)_{MN}=det(g)_{MN}+
det(\lambda F)_{MN}+\cdots$, where the ellipses stands for various even powers of $F$. Upon restricting to $3+1$ dimensional spacetime, the $det(F)_{MN}\propto F\w F$ and this term is absent in the action of the BI matter.}, but in a specific situation they can coincide. This happen only when the U(1) field strength has got  one non-vanishing component. Let us illustrate this point by considering two non-vanishing component of the field strength, $
F=A'_t(r) dr\w dt+B dx\w dy$, 
 in $d+1$ dimensional spacetime,  $ds^2_{d+1}=-g_{tt}(r)dt^2+g_{xx}(r)(dx^2+dy^2)+g_{xx}(r)dz^2_i+g_{rr}(r)dr^2$. 

On computing the BI and the  DBI matter 
\bea
S_{BI}&=&\int \sqrt{g_{tt}g_{rr}}g^{\f{d-1}{2}}_{xx}
\sqrt{1-2\alpha\f{A'^2_t}{g_{tt}g_{rr}}+2\alpha \f{B^2}{g^2_{xx}}},\nn S_{DBI}&=&\int\sqrt{g_{tt}g_{rr}}
g^{\f{d-1}{2}}_{xx}\sqrt{1-\f{\lambda^2A'^2_t}{g_{tt}g_{rr}}+\f{\lambda^2B^2}{g^2_{xx}}-\f{\lambda^4 B^2 A'^2_t}{g_{tt}g_{rr}g^2_{xx}}}.
\eea
So, there follows that either for zero magnetic field or for zero charge density, $A'_t=0$, both BI and DBI gives the same action for $2\alpha=\lambda^2$. Hence, its only the electrically charged black hole solution is same as found in \cite{Dey:2004yt}, but not the dyonic solution. This we illustrate by finding the exact  solution to the Einstein-cosmological constant-BI action in Appendix B.

\subsection{Solution with non-trivial scalar field}

In this section, we shall find the solution at UV with a non-trivial dilaton profile. In order to do so, we shall write down the equation of motion of the metric in the following form 
\bea
&&R_{MN}-\f{1}{2}g_{MN} R-\f{1}{2}\p_M\phi\p_N \phi+\f{1}{2}g_{MN}\left(2\Lambda+V(\phi)+\f{1}{2}g^{KL} \p_K\phi\p_L \phi\right)+
\f{T_b}{4} Z_1(\phi)Z_2(\phi)\nn&&\times\f{\sqrt{-det(Z_2g+\lambda F)_{ab}}}{\sqrt{-g}}\bigg[\bigg(g~ Z_2(\phi)+\lambda F\bigg)^{-1KL}+{\bigg(g~ Z_2(\phi)+\lambda F\bigg)}^{-1LK} \bigg]g_{KM}g_{NL}=0
\eea

and the rest of the equations of motion are as written down in eq(\ref{gauge_field_eom}) and eq(\ref{scalar_eom}). We shall consider the ansatz as well as the functions as written down in eq(\ref{ansatz_solution}), eq(\ref{def_alpha_beta}) and eq(\ref{potential}), respectively.
Using this ansatz  for the metric,  the dilaton and the solution for the gauge field is as written down in eq(\ref{sol_gauge_field}),  we find the equations of motion of the metric component  reduces to 
\bea\label{G_equals_T}
&&R_{tt}+\f{1}{2}g_{tt} R-\f{1}{2}g_{tt}\left(2\Lambda+V(\phi)+\f{1}{2}g^{KL} \p_K\phi\p_L \phi\right)-\f{T_b}{2} \f{Z_2 g_{tt}}{g^{(d-1)/2}_{xx}}\sqrt{\rho^2+Z^2_1Z^{d-1}_2g^{d-1}_{xx}}=0,\nn
&&R_{xx}-\f{1}{2}g_{xx} R+\f{1}{2}g_{xx}\left(2\Lambda+V(\phi)+\f{1}{2}g^{KL} \p_K\phi\p_L \phi\right)+\f{T_b}{2} \f{Z^2_1Z^d_2 g^{(d+1)/2}_{xx}}{\sqrt{\rho^2+Z^2_1Z^{d-1}_2g^{d-1}_{xx}}}=0,\nn
&&R_{rr}-\f{1}{2}g_{rr} R-\f{\phi\rq{}^2}{2}+\f{1}{2}g_{rr}\left(2\Lambda+V(\phi)+\f{1}{2}g^{KL} \p_K\phi\p_L \phi\right)+\nn&&~~~~~~~~~~~~~~~~~~~~~~~~~~~~~~~~~~~~~\f{T_b}{2} \f{Z_2 g_{rr}}{g^{(d-1)/2}_{xx}}\sqrt{\rho^2+Z^2_1Z^{d-1}_2g^{d-1}_{xx}}=0,
\eea

Now using the first and the last equations of eq(\ref{G_equals_T}), we find
\be\label{relation_Rrr_Rtt}
R_{rr}+\f{g_{rr}}{g_{tt}}R_{tt}=\f{\phi\rq{}^2}{2}.
\ee

In what follows, we shall use eq(\ref{relation_Rrr_Rtt}), the second and the last equation of  eq(\ref{G_equals_T}) to solve the equations of motion of the metric. Let us take the following  explicit (asymptotically non-AdS spacetime)  ansatz to the metric and  dilaton  
\be\label{geometry_dilaton_dbi}
ds^2_{d+1}=-r^2f(r)dt^2+\f{e^{2\Theta(r)}dr^2}{r^2f(r)}+r^2 dx^idx_i,\quad 
\phi= \delta~ Log~r,
\ee
where $\delta$ is a constant\footnote{Such a form  of the solution is adopted for a $2+1$ dimensional Einstein-BI-dilaton system in \cite{Yamazaki:2001ue} following  \cite{Clement:2000ue}.}.  Upon substituting such a form of the metric  into eq(\ref{relation_Rrr_Rtt})
gives\footnote{Had we taken an asymptotically AdS spacetime anstz to the geometry then  eq(\ref{relation_theta_phi})  would have given a constant dilatonic solution, which we derived in the previous subsection.}
\be\label{relation_theta_phi}
\phi\rq{}^2=\f{2(d-1)}{r}\Theta\rq{}.
\ee
For a logarithmic dilaton, it means the function $\Theta$ should better be logarithmic, $\Theta(r)=\delta_1 ~Log~r$. Moreover, one can show that the only relevant equation  that we need to solve is a first order differential equation, which follows from the  last equation of eq(\ref{G_equals_T}) for the following choice of the constants.

\be
\alpha=-\sqrt{\f{(d-1)\delta_1}{2}},\quad \beta=m_2=-\sqrt{\f{2\delta_1}{(d-1)}},\quad \delta=\sqrt{2(d-1)\delta_1},\quad \Lambda=0,\quad V=m_1~ r^{-2\delta_1}. 
\ee
In which case the function, $f(r)$, satisfies the following differential equation 
\be
rf\rq{}(r)+(d-\delta_1)f(r)+\f{m_1}{(d-1)}+\f{T_b}{(d-1)}r^{1-d}\sqrt{\rho^2+r^{2(d-1)}}=0.
\ee

On solving this differntial equation, we find the solution as
\bea\label{f_dilaton_dbi}
f(r)&=&\f{c_1}{r^{d-\delta_1}}-\f{m_1}{(d-\delta_1)(d-1)}-\nn&&\f{T_b}{(d-1)(\delta_1-1)}\rho r^{1-d}{}_2F_1\left[-\f{1}{2},\f{1-\delta_1}{2(d-1)},\f{2d-1-\delta_1}{2(d-1)},-\f{r^{2(d-1)}}{\rho^2} \right],\eea

where $c_1$ is the constant of integration and can identified with the mass density of the solution. The gauge field is 
\be
\lambda A\rq{}_t(r)=\f{\rho}{\sqrt{\rho^2+r^{2(d-1)}}}r^{-\delta_1}.
\ee

In the $\delta_1\ra 0$ limit, this solution reduces to the solution as written in eq(\ref{sol_f_without_scalar}) provided we set $m_1=2\Lambda$. 

The Hawking temperature and the Bekenstein-Hawking entropy density of such  black hole becomes
\bea
T_H&=&-\f{r^{1-2\delta_1}_h}{4\pi(d-1)}\left[m_1+T_b r^{1-d}_h\sqrt{\rho^2+r^{2(d-1)}_h} \right],\nn
s&=&\f{2\pi}{\kappa^2}~r^{d-1}_h,
\eea
where $r_h$ is the horizon and determined from the zero of the function $f(r_h)=0$. The specific heat becomes
\be
C_v=\f{2(d-1)\pi r^d_h \sqrt{\rho^2+r^{2(d-1)}_h}[m_1 r^d_h+T_b r_h \sqrt{\rho^2+r^{2(d-1)}_h} ] }
{\kappa^2 \left[T_b r^{2d}_h(1-2\delta_1)+T_b r^2_h\rho^2 (2-d-2\delta_1)+m_1 r^{d+1}_h (1-2\delta_1) \sqrt{\rho^2+r^{2(d-1)}_h}  \right]}.
\ee

The  positivity of the specific heat forces some choice of the constant, $\delta_1$. In order to see it, let us set $r_h=1$, for simplicity.
On expanding the specific heat for small charge density
\be
C_v=\f{2(d-1)\pi}{(1-2\delta_1)\kappa^2}+\f{2(d-1)^2 T_b\pi }{(T_b+m_1)(1-2\delta_1)^2\kappa^2}\rho^2+{\cal O}(\rho^4).
 \ee 

It is highly plausible to set  a restriction on the parameter  as $\delta_1 < 1/2$, whereas in the high charge density limit, the specific heat, $C_v=\f{2\pi(d-1)r^{d-1}_h}{\kappa^2(2-d-2\delta_1)}+{\cal O}\left(\rho^{-2}\right)$. Once again demanding the positivity of the specific heat requires us to set the constraint as $\delta_1 < 1-d/2$. So, it follows that for high $d$, the constant $\delta_1$ can become negative.

\subsection{Dyonic dilaton solution}

In this case with a constant magnetic field and the  gauge field strength as written in eq(\ref{dyonic_field_strength}) gives the  solution for the field strength as written down in eq(\ref{dyonic_gauge_field_sol}). The ansatz for the geometry is assumed to be of the form as written in eq(\ref{geometry_dilaton_dbi}).
Without giving the details for the metric and the dilaton equations of motion, we simply   
 give the solution
\bea\label{dyonic_dilaton_dbi}
ds^2_{3+1}&=&-r^2 f(r) dt^2+r^2(dx^2+dy^2)+\f{dr^2}{f(r)},\quad \phi=2~ Log~r,\quad \Theta=Log~r,\nn
V&=&\f{m_1}{r^2},\quad \alpha=-\f{1}{1+\lambda^2 B^2},\quad \beta=m_2=-1,\quad \Lambda=0,\nn
f(r)&=&\f{c_1}{r^2}-\f{m_1}{4}+\f{T_b}{4r^2\alpha}\sqrt{\rho^2+\f{(1+\lambda^2 B^2)}{r^{4\alpha}}}-\f{T_b\rho}{4r^2\alpha} Tanh^{-1}\left(\sqrt{1+\f{(1+\lambda^2 B^2)}{\rho^2r^{4\alpha}}} \right),\nn
\lambda A\rq{}_t(r)&=&\f{\rho}{r\sqrt{\rho^2+(1+\lambda^2 B^2)r^{-4\alpha}}}.
\eea

In the limit of vanishing magnetic field, $B\ra 0$, we reproduce the solution for the charged black hole as written in the previous section in $3+1$ dimensional spaetime in the limit  of $\delta_1=1$.

\section{Application: DC conductivity}

As an application of the solutions found in the previous sections, 
we shall study various properties of  the action as written in eq(\ref{eh_dbi_dilaton}). To begin with, we shall calculate the dc conductivity. To do the computations, we shall use the flow equation technique of \cite{Iqbal:2008by}, which is done in \cite{Lee:2011qu} and \cite{Pal:2012gr} for the DBI system. In order to do the computations, let us first fluctuate both the metric and the gauge field components. For simplicity, we shall restrict the fluctuation to $g_{tx}$ and $A_x$ component.  Also $g_{tx}$ and $A_x$ and are assumed to be functions of time, $t$ and $r$. The time dependence of the fluctuating fields comes e.g. in the metric fluctuation as $g_{tx}(r) e^{-i\omega t}$.  In what follows, we shall not be computing the conductivity for those solutions which has a  non-zero magnetic field. Doing the above mentioned fluctuations\footnote{We shall work in the radial gauge: $A_r=0,~g_{Mr}=0$. It follows that  even after fixing the gauge choice there exists some residual symmetry that of the U(1) gauge invariance and the diffeomorphism invariance. In cases where there exists metric, U(1) gauge field and a scalar field,  the precise form of the residual symmetry is written down in eq(79)-(81) of \cite{Anantua:2012nj}. In our case, we can still take that   diffeomorphism invariance because for small value of the field strength, i.e., in the dilute limit, the DBI action reduces to the Maxwell action.
Moreover, when the momentum is along the x-direction, there exists two different kind of modes depending on the $y\ra -y$. Here we are considering the fluctuations that are part of the longitudinal mode  \cite{Anantua:2012nj}. In doing the analysis, we have set some of the fluctuating fields to zero {\it a la} \cite{Hartnoll:2008kx}. In fact, it is very interesting to do  the detailed analysis by keeping  all the fluctuating degrees of freedom which we defer for future.  }, we find that the $x-r$ component of the equation of motion as written in eq(\ref{metric_eom}) gives the following relation between the metric fluctuation and the gauge field fluctuation
\be\label{relation_metric_gauge_field_fluctuation}
\sqrt{g_{tt}g_{rr}Z^2_2-\lambda^2 A'^2_t}(g'_{xx}g_{tx}-g_{xx}g'_{tx})+\lambda^2 T_b Z_1 Z^{\f{d-1}{2}}_2 g_{xx}A_x A'_t \sqrt{g_{tt}g_{rr}}=0,
\ee
where prime denotes derivative with respect to $r$ and we have done the Fourier transformation with respect to $e^{-i\omega t}$, means we have set the momentum to zero. Also, in doing the computation, we have used the following result to the Ricci tensor $R_{xr}=\f{i\omega}{2g_{tt}g_{xx}}(g'_{xx}g_{tx}-g_{xx}g'_{tx})$.  

Let us expand the gauge field part of the action as written in eq(\ref{eh_dbi_dilaton}) to quadratic order in the gauge field, $A_x$, using eq(\ref{relation_metric_gauge_field_fluctuation})  results in
\be\label{action_A_x}
S^{(2)}_A=-\f{\lambda^2T_b}{4\kappa^2}\int \f{\sqrt{\rho^2+Z^2_1 Z^{d-1}_2g^{d-1}_{xx}}}{Z_2 g_{xx}\sqrt{g_{tt}g_{rr}}}\left[ g_{tt}A'^2_x-A^2_x\left(\omega^2 g_{rr}+\f{2\lambda^2 T_b A'^2_t\sqrt{\rho^2+Z^2_1 Z^{d-1}_2g^{d-1}_{xx}}}{Z_2 g^{\f{d-1}{2}}_{xx}} \right)\right].
\ee

The equation of motion that follows from it takes the following form
\be
\p_r\left[  \f{\sqrt{\rho^2+Z^2_1 Z^{d-1}_2g^{d-1}_{xx}}}{Z_2 g_{xx}\sqrt{g_{tt}g_{rr}}}g_{tt}A'_x \right]+ \f{\sqrt{\rho^2+Z^2_1 Z^{d-1}_2g^{d-1}_{xx}}}{Z_2 g_{xx}\sqrt{g_{tt}g_{rr}}} \left(\omega^2 g_{rr}+\f{2\lambda^2 T_b A'^2_t\sqrt{\rho^2+Z^2_1 Z^{d-1}_2g^{d-1}_{xx}}}{Z_2 g^{\f{d-1}{2}}_{xx}} \right)A_x=0.
\ee

Let us compute the current  at some choice of the radial coordinate,  $r=r_c$, from eq(\ref{action_A_x})
\be\label{current}
J^x(r_c)=-\f{\lambda^2T_b}{2\kappa^2}\Bigg[  \f{\sqrt{\rho^2+Z^2_1 Z^{d-1}_2g^{d-1}_{xx}}}{Z_2 g_{xx}\sqrt{g_{tt}g_{rr}}}g_{tt}A'_x \Bigg]_{r=r_c}.
\ee

Now assuming that  the Ohm's law holds at any choice of the the radial slice, $r_c$, gives 
\be\label{ohms_law}
J^x(r_c)=\sigma^{xx}(r_c,\omega)E_x(r_c)=\sigma^{xx}(r_c,\omega) i\omega A_{x}(r_c),
\ee
where in the second equality we have expressed all the quantities in the Fourier space. 
%The action for the fluctuating gauge field as written  in eq(\ref{action_A_x})  looks very similar to the action for the scalar field in a different spacetime. For the scalar field the retarded correlator at the boundary is given in eq(29) of \cite{il}. 
In order to see such a form of the Ohm's law, we assume that the  retarded correlator at any radial slice is given, by generalizing eq(29) of \cite{Iqbal:2008by}
\be
G_R(r_c,k_{\mu})=-\f{\Pi(r_c,k_{\mu})}{A_x(r_c,k_{\mu})},
\ee
where $\Pi(r_c,k_{\mu})$ is the momentum associated to the field $A_x$, evaluated at $r_c$. In this case, the transport quantity, which is the conductivity,  at $r_c$ is related to the  retarded correlator as $\sigma(r_c,k_{\mu} )=i G_R(r_c,k_{\mu})/\omega=-i\Pi(r_c,k_{\mu})/A_x(r_c,k_{\mu})$. Note that the current in eq(\ref{current}) is nothing but the momentum associated to $A_x$, hence there follows the Ohm's law at  slice $r_c$, eq(\ref{ohms_law}).

Let us evaluate the flow equation of the conductivity as we change the slice from $r_c$ to $r_c+\delta r_c$ in the limit $\delta r_c \ra 0$. In which case, the resulting flow equation becomes
\be\label{flow_cond}
\p_{ r_c}\sigma^{xx}( r_c,\omega )=-i\omega \sqrt{\f{g_{rr}( r_c)}{g_{tt}( r_c)}}\left[\Sigma_A( r_c)-\f{(\sigma^{xx}(r_c,\omega))^2}{\Sigma_A( r_c)} +\f{4\lambda^2 T^2_b \rho^2}{4\kappa^2}\f{Z_2( r_c) g_{tt}( r_c)}{g^{\f{d+1}{2}}_{xx}( r_c)}\right],
\ee
where $\Sigma_A=2\f{\lambda^2 T_b}{4\kappa^2} \f{\sqrt{\rho^2+Z^2_1z^{d-1}_2g^{d-1}_{xx}}}{Z_2 g_{xx}}$. At the horizon, the time component of the metric vanishes, $g_{tt}(r_h)=0$, which means as  we take the limit $r_c\ra r_h$, we  need to impose a regularity condition on the conductivity at the horizon and the condition reads
\be\label{cond_horizon}
\sigma^{xx}(r_h)=\Sigma_A(r_h)= 2\f{\lambda^2 T_b}{4\kappa^2}\Bigg[ \f{\sqrt{\rho^2+Z^2_1Z^{d-1}_2g^{d-1}_{xx}}}{Z_2 g_{xx}}\Bigg]_{r_h}.
\ee  

It is interesting to note that at the horizon the in-falling boundary condition for the gauge field, $A_x$,  follows naturally combining the  form of the conductivity at the horizon eq(\ref{cond_horizon})  and the Ohm's law at the horizon,  i.e., $J^x(r_h)=\sigma^{xx}(r_h)i\omega A_x(r_h)$. In order to see it, let us  use eq(\ref{current}) and eq(\ref{cond_horizon}) in the Ohm's law. Then it follows that  the derivative of the gauge field is related to the gauge field at the horizon as 
\be
A'_x(r_h)=-i\omega \Bigg[\sqrt{\f{g_{rr}}{g_{tt}}}A_x\Bigg]_{r_h}.
\ee

Integrating this gives us the desired in-falling  form of the gauge field at the horizon, namely
\be
A_x(r_h)=e^{-i\omega \int^{r_h}dr\sqrt{\f{g_{rr}}{g_{tt}}}}.
\ee 

It is very easy to convince that in the zero frequency limit, i.e., the DC conductivity remains same over  the entire range of the radial coordinate, which suggests that it does not run.  Now given the form of the DC conductivity as in eq(\ref{cond_horizon}), there follows the following temperature dependence for different solutions
\be
\sigma^{xx}=\f{\lambda^2 T_b}{2\kappa^2} \sqrt{1+\rho^2}\times 
   \left\{
  \begin{array}{l l}
 (\f{z+2}{4\pi})^{(2/z)}~
    T^{-\f{2}{z}}_H & \quad \textrm{for eq(\ref{bh_dbi_3+1_d_IR}) }\\
  (\f{z+2}{4\pi})^{4/z}~ T^{-4/z}_H   & \quad \textrm{for eq(\ref{lifshitz_sol})}.\\
 \end{array} \right.
\ee

Let us recall that the longitudinal conductivity for the non-Fermi liquid (NFL) state of the matter has the inverse temperature dependence. Now at IR, we do generate such a behavior of the conductivity, if we choose  the dynamical exponent $z=2$ for the solution   eq(\ref{bh_dbi_3+1_d_IR}) whereas  $z=4$ for the solution written in  eq(\ref{lifshitz_sol}).  Such a behavior of  the longitudinal conductivity  has been found in \cite{Hartnoll:2009ns} for $z=2$, in the probe brane approximation.  However, for $z=2$, we find that the solution as written in  eq(\ref{lifshitz_sol}) has the longitudinal conductivity, $\sigma^{xx}\sim T^{-2}_H$ and that of eq(\ref{bh_dbi_3+1_d_IR}) has the longitudinal conductivity, $\sigma^{xx}\sim T^{-1}_H$.
This implies that as far as the conductivity is concerned, by tuning the parameters like $\alpha$ and $\beta$, we can describe either the FL or the NFL state in a $2+1$ dimensional field theory. Such a crossover was found  in  a $3+1$ dimensional field theory   using a magnetic field in \cite{Pal:2012gr}.

\section{Entanglement entropy}

The entanglement entropy of a $d$ dimensional field theory  or a $d+1$ dimensional gravitational system  is determined by finding 
a $d-1$ dimensional minimal spacelike hypersurface, $\gamma_A$,      that extremizes the area of the hypersurface \cite{Ryu:2006bv}.  The explicit formula for the  entanglement entropy as suggested in  \cite{Ryu:2006bv} takes the following form:  $S_A=\f{{\rm Area ~of}~ \gamma_a}{4G_N}$, where $G_N$ is the Newton's constant in $d+1$ dimensional gravitational system.  Various aspects of the  entanglement entropy is further studied e.g.,  in \cite{Li:2009pf}-\cite{Kulaxizi:2012gy}.

In order to do the computation, let us move to a coordinate system $u$ for which the boundary is at $u=0$ and assume that the bulk spacetime takes the  following form
\be
ds^2_{d+1}=-g_{tt}(u)dt^2+g_{xx}(u)dx^2_i+g_{uu}(u)du^2.
\ee
In which case the geometry of the  $d-1$ dimensional  hypersurface takes the following form
\be
ds^2_{d-1}=\Bigg[g_{uu}(u)\Bigg(\f{du}{dx_1}\Bigg)^2+g_{xx}(u)\Bigg]dx^2_1+g_{xx}(u)(dx^2_2+\cdots+dx^2_{d-1}),
\ee 
where the precise nature of the hypersurface is determined by the function $u(x_1)$.
On computing the area of the hypersurface,  $\gamma_A$, we find
\be\label{area_genric}
{\cal A} (\gamma_A)= \int dx_2\cdots\int dx_{d-1}\int du~ g^{\f{d-2}{2}}_{xx}~\sqrt{g_{uu}+g_{xx} (dx_1/du)^2}.
\ee

In order to carry out the integral of $x_2$ to $x_{d-1}$, let us assume, for simplicity, the hypersurface has the shape of a strip. In which case, we assume that $-\ell \leq x_1 \leq \ell$ and $0\leq (x_2,\cdots, x_{d-1})\leq L$. Finally, performing the above mentioned  integrals result in 
\be\label{area_ads2}
{\cal A} (\gamma_A)= L^{d-2}\int^{\ell}_{-\ell} dx_1~ g^{\f{d-2}{2}}_{xx}~\sqrt{g_{xx}+g_{uu} (du/dx_1)^2}.
\ee

Extremizing the surface area gives the following solution to the function $u(x_1)$
\be\label{fun_u_x_1}
\f{du}{dx_1}=\f{\sqrt{g^d_{xx}(u)-g_{xx}(u) g^{d-1}_{xx}(u_{\star})}}{g^{\f{d-1}{2}}_{xx}(u_{\star})\sqrt{g_{uu}(u)}},
\ee 
where the turning point,  $u_{\star}$,  is determined as the point where $(\f{dx_1}{du})_{u_{\star}}$ diverges. The length along $x_1$ is
\be\label{turning_pt}
\ell=\int^{u_{\star}}_0 du\f{g^{\f{d-1}{2}}_{xx}(u_{\star})\sqrt{g_{uu}(u)}}{\sqrt{g^d_{xx}(u)-g_{xx}(u) g^{d-1}_{xx}(u_{\star})}}=g^{\f{d-1}{2}}_{xx}(u_{\star})\int^{u_{\star}}_0 du \sqrt{\f{g_{uu}(u)}{g^{d}_{xx}(u)}}\f{1}{\sqrt{1-\f{ g^{d-1}_{xx}(u_{\star})}{g^{d-1}_{xx}(u)}}}
\ee

Finally, substituting this form of the function, $u(x_1)$, from eq(\ref{fun_u_x_1}) into the area gives
\be\label{area_ent}
{\cal A} (\gamma_A)= L^{d-2} \int^{u_{\star}}_{\epsilon} du~ \f{g^{d-1}_{xx}(u)\sqrt{g_{uu}(u)}}{\sqrt{g^d_{xx}(u)-g_{xx}(u) g^{d-1}_{xx}(u_{\star})}}= L^{d-2}
\int^{u_{\star}}_{\epsilon} du~ \f{\sqrt{g^{d-2}_{xx}(u)g_{uu}(u)}}{\sqrt{1-\f{ g^{d-1}_{xx}(u_{\star})}{g^{d-1}_{xx}(u)}}},
\ee
where $\epsilon$ is the UV-cutoff which will regulate the presence of the divergence while approaching the boundary, i.e.,  taking the 
$u\ra 0$ limit. Generically, to perform the $u$ integration  is not easy. So, to evaluate the area, let us use   eq(\ref{area_genric}), instead. After doing the integrals of $x_2$ to $x_{d-1}$, we obtain  
\be
{\cal A} (\gamma_A)= L^{d-2}\int^{u_{\star}}_{\epsilon} du~ g^{\f{d-2}{2}}_{xx}~\sqrt{g_{xx}(dx_1/du)^2+g_{uu} }
\ee

and assume that close to the boundary, the velocity, $dx_1/du\ra 0$, is small. It means we can do a Taylor series expansion there. The leading order term gives
\be\label{area_uv}
{\cal A}_{UV} (\gamma_A)\simeq  L^{d-2} \int_{\epsilon}   du~\left[ g^{\f{d-2}{2}}_{xx} \sqrt{g_{uu}}+{\cal O}(dx_1/du)^2\right],
\ee
whereas away from the boundary, $u\ra u_{\star}$, the  velocity, $dx_1/du\ra \infty$. This diverging nature essentially follows from eq(\ref{fun_u_x_1}). In which case, we can approximate the area as
\bea\label{area_ir}
{\cal A}_{IR} (\gamma_A)&\simeq&  L^{d-2} \int^{u_{\star}} du  ~\left[g^{\f{d-1}{2}}_{xx} ~\f{dx_1}{du}+{\cal O}(du/dx_1)^2\right]\nn
&=& L^{d-2} \int^{u_{\star}}_{u_F} du  ~\f{g^{\f{d-1}{2}}_{xx}(u_{\star})\sqrt{g_{uu}(u)}}{\sqrt{g_{xx}(u)}}+\cdots
\eea

In going to the second line we have made another assumption: $g_{xx}(u_{\star})/g_{xx}(u)<<1$ in the range $ u_F\leq u < u_{\star}$.
In which case, the velocity can be approximated as
\be
\f{dx_1}{du}\approx \f{g^{\f{d-1}{2}}_{xx}(u_{\star})\sqrt{g_{uu}(u)}}{g^{d/2}_{xx}(u)}.
\ee
 
 Now let us ask the question:  Under what condition, we  get the desired area law and the log violation of it? In order to answer both the questions, we must impose the following conditions
\be
g^{d-2}_{xx}g_{uu}\sim u^{2(1-d)},\quad {\rm and} \quad \f{g_{uu}}{g_{xx}}\sim 1/u^2.
\ee

On solving these conditions, there follows 
\be\label{metric_log_violation_area}
g_{xx}(u)\sim R^2~u^{-\f{2(d-2)}{d-1}},\quad g_{uu}(u)\sim R^2~ u^{-\f{2(2d-3)}{d-1}}.
\ee

Note that the overall constant factor is not  determined.  
It is easy to see that our assumption,  $g_{xx}(u_{\star})/g_{xx}(u_F)<<1$, holds for  $ u_F <  u_{\star}$. The parameter $u_F$ in the boundary field theory is interpreted as the scale of the Fermi surface. Finally, the area at the UV and IR takes the following form
\be
{\cal A}_{UV}\simeq \f{R^{d-1}}{d-2}\left(\f{L}{\epsilon}\right)^{d-2},\quad {\cal A}_{IR}\simeq  R^{d-1}\left(\f{L}{u_{\star}}\right)^{d-2}~Log\left( \f{u_{\star}}{u_F}\right).
\ee
%In \cite{Ogawa:2011bz}, it is suggested that 
%the parameter $u_F$ should be smaller than the size $\ell$. 
%\cite{Ogawa:2011bz}. Let us consider the following form of 

%\be
%g_{xx}(u)=\Sigma_x(u_{\star}) u^{-\f{2(d-2)}{d-1}},\quad g_{uu}(u)= \Sigma_u(u_{\star}) u^{-\f{2(2d-3)}{d-1}},
%\ee
%where $\Sigma_x(u_{\star})$ and $\Sigma_u(u_{\star})$ are constants. In order to fix these constants, we must impose the condition that the area of the hypersurface at the boundary, i.e., eq(\ref{area_uv}) should not be a function of $u_{\star}$. Moreover,  the length $\ell\sim u_{\star}$. This fixes the constants as
%\be
%\Sigma_x(u_{\star})\sim u^{\f{2}{1-d}}_{\star},\quad \Sigma_u(u_{\star})\sim u^{\f{2(d-2)}{d-1}}_{\star}.
 %\ee 

Since the $g_{tt}(u)$ component of the metric is not fixed by the entanglement entropy, it  means the bulk geometry that gives the log violation of the area law, within the assumptions as mentioned above, should have the following form 
\be
ds^2_{d+1}=R^2\left[-g_{tt}(u)dt^2+\Sigma_x(u_{\star})u^{-\f{2(d-2)}{d-1}}dx^2_i+\Sigma_u(u_{\star})u^{-\f{2(2d-3)}{d-1}}du^2\right],
\ee
where $\Sigma_x$ and $\Sigma_u$ are constants
\footnote{For $\Sigma_x(u_{\star})\sim u^{\f{2}{1-d}}_{\star},\quad \Sigma_u(u_{\star})\sim u^{\f{2(d-2)}{d-1}}_{\star}$, it just simply follows that  $\ell\sim u_{\star}$. Note that the  $\Sigma$\rq{}s can be  absorbed  into $t,~x$ and $R$, in which case   there exists a caveat. 
Even though the geometry as written in eq(\ref{metric_log_violation_area}) gives us the necessary area law at the UV and the logarithmic violation of it at IR, but   the length of the strip along $x_1$ direction is   independent of the location of $u_{\star}$ or $u_F$. In order to see it, 
let us use the form  of  $g_{xx}$ and $g_{uu}$ in eq(\ref{turning_pt}) and after a change of variable, the integral can be re-written as 
\be
\ell=\int^1_0 dt\f{t^{d-3}}{\sqrt{1-t^{2(d-2)}}}.
\ee
%The  dimensionless feature of  $\ell$   is expected because $x_i$\rq{}s are dimensionless.
}.
Using a different coordinate system $u=1/r^2$ the geometry becomes
\be
ds^2=R^2\left[-g_{tt}(r)dt^2+\Sigma_x(u_{\star})r^{\f{4(d-2)}{d-1}}dx^2_i+4\Sigma_u(u_{\star}) r^{\f{2(d-3)}{d-1}}dr^2 \right],
\ee

which for $d=3$ with a  re-definition of  the time $t$, spatial coordinate $x_i$ and $R$ gives the geometry as
\be
ds^2_{3+1}=R^2\left[-g_{tt}(r)dt^2+r^2(dx^2+dy^2)+ dr^2 \right].
\ee

For a specific choice of the $g_{tt}(r)$ component of the metric and setting $R=1$ gives the zero temperature limit of the geometry as written in eq(\ref{bh_dbi_3+1_d_IR}).\\

\underline{A comment:}\\

The result presented in  \cite{Huijse:2011ef},  namely,  $\theta=d-2$, which gives the logarithmic violation of the entanglement entropy  can  
easily be derived from  eq(\ref{area_ent}) using the geometry as written down in eq(\ref{most_general_geometry_with_all_possibilities}), for $\delta=1$. Upon working in a coordinate system for which the boundary is at $r=\infty$, the entanglement entropy for a strip 
takes the following form
\be
{\cal A} (\gamma_A)= L^{d-2} \int^{\Lambda}_{r_{\star}} dr~ \f{g^{d-1}_{xx}(r)\sqrt{g_{rr}(r)}}{\sqrt{g^d_{xx}(r)-g_{xx}(r) g^{d-1}_{xx}(r_{\star})}}= L^{d-2}
\int^{\Lambda}_{r_{\star}} dr~ \f{\sqrt{g^{d-2}_{xx}(r)g_{rr}(r)}}{\sqrt{1-\f{ g^{d-1}_{xx}(r_{\star})}{g^{d-1}_{xx}(r)}}},
\ee
where $\Lambda$ is the UV-cutoff which will regulate the presence of the divergence, while taking the $r\ra \infty$ 
limit. Essentially, the form of the expression of the entanglement entropy for both the $r$ and $u$ coordinate system  remains the same except the limits of the integral.

If we impose the condition that there should be a logarithmic term in the entanglement entropy means we need to set
%\footnote{Had we used   eq(\ref{area_ent}) and set $g^{d-2}_{xx} g_{uu}\sim 1/u^2$ then we would have obtained the log term but with a wrong sign.} 
\be\label{approximate_log_cond}
g^{d-2}_{xx} g_{rr}\sim 1/r^2.
\ee

Upon using eq(\ref{most_general_geometry_with_all_possibilities}),  there  follows the condition
\be\label{theta_delta_relation}
\delta(d-2)=\gamma(d-1)\equiv\theta,
\ee

which for  $\delta=1$, reproduces  the result of  \cite{Huijse:2011ef}. Note that  this method does not work for $\delta=0$ case. Because for $\delta=0$, we need to set $\gamma=0$,  which implies that   the geometry should be  $AdS_2\otimes R^{d-1}$, and it is known from the previous studies, this geometry  does not give any logarithmic violation of the
entanglement entropy  \cite{Swingle:2009wc}.  Let us  illustrate this case in detail. 
%The basic reason of the failure of the formula eq(\ref{theta_delta_relation}) for the $\delta=0$ and $\gamma=0$ case is due to the breakdown of an implicit assumption. The assumption is that for $r_F> r> r_{\star} $, the quantity $g_{xx}(r_{\star}) / g_{xx}(r_F)=(r_{\star}/r_F)^{2(\delta-\gamma)} <1$, where we have used eq(\ref{most_general_geometry_with_all_possibilities}). But this assumption is broken for  $\delta=0=\gamma$, as $g_{xx}(r_{\star}) = g_{xx}(r_F)$.

The failure of the applicability of the formula
eq(\ref{approximate_log_cond}) for  $\delta=0$ and $\gamma=0$ can be seen  as follows. 
%This is essentially due to the failure of the formula for the inverse velocity as written in eq(\ref{fun_u_x_1}).
 Recall that for $AdS_2\otimes R^{d-1}$ spacetime the metric component $g_{xx}$ is constant, which means the turning point $u_{\star}$ has to be determined, carefully. 
In which case, the velocity and the length of the strip along $x_1$ direction becomes
\be
\f{dx_1}{du}=\f{c_0\sqrt{g_{uu}(u)}}{\sqrt{g^d_{xx}-g_{xx}c^2_0}},\quad \ell=\f{c_0}{\sqrt{g^d_{xx}-g_{xx}c^2_0}}\int^{u_{\star}}_0du  \sqrt{g_{uu}(u)},
\ee
where the constant $c_0$ is  fixed by requiring the condition that as $u\ra u_{\star}$, the velocity diverges, i.e., $(\f{dx_1}{du})_{u_{\star}}\ra\infty$.
Finally, the area of the hypersurface becomes
\be
{\cal A}=L^{d-2}  \f{g^{d-1}_{xx}}{\sqrt{g^d_{xx}-g_{xx}c^2_0}} \int^{u_{\star}}_0du  \sqrt{g_{uu}(u)}= \f{ L^{d-2}~ g^{d-1}_{xx}~\ell}{c_0}.
\ee

The velocity diverges only when $c^2_0=g^{d-1}_{xx}$, which means the area of the hypersurfce\footnote{Since the metric component $g_{xx}$ is constant means it can absorbed into the coordinate $x_i$, in which case the area simply becomes ${\cal A}=L^{d-2}~\ell$.} becomes
\be
{\cal A}=L^{d-2}~\ell~g^{\f{d-1}{2}}_{xx}.
\ee

\section{Conclusion}

In this paper, we have obtained the geometry both at IR and UV,    by considering the back reaction of a space filling D-brane in the presence of a scalar field, i.e., 
new solutions to Einstein-DBI-dilaton system. 

At IR, we suggest a  form of the entropy density, which has the following temperature dependence 
\be
s\sim T^{\f{\delta(d-1)-\theta}{z}}_H.
\ee

This particular form of the entropy density  differs from that given\footnote{Which is a special case of this formula.  Note that
the above form of the entropy density is backed up by finding the explicit solutions in generic spacetime dimensions.} in \cite{Huijse:2011ef}  essentially because of the way the spatial field theoretic  directions scale under scaling transformation. It is described by a parameter, $\delta$, which can take only two values $0$ or $1$. In \cite{Huijse:2011ef}  only the linearly scaling behavior of the spatial field theoretic directions were considered. Here, we have considered the other case, namely, when it does not scale\footnote{ This  case is also discussed in \cite{Hartnoll:2012wm} and \cite{Anantua:2012nj}.}. The null energy condition suggests that the scaling violating exponent, $\theta$, should be negative\footnote{  Which essentially mean this formula for the entropy density does not work when both $\delta$ and $\theta$ vanishes, which is the case for $AdS_2\otimes R^{d-1}$ because this particular case is subtle.} and the dynamical exponent, $z$,  should  be positive.

The most notable solution for $\theta=0$ and $\delta=1$ shows the Fermi-like liquid  behavior in   $3+1$ dimensional gravitational description. In this  case we generate the black hole solution for  Lifshitz spacetime. For a specific choice of the dynamical exponent, $z=2$, the longitudinal conductivity and the specific heat goes as inverse quadratic and linear in temperature, respectively. 
Even though the geometry and the  U(1) gauge field is scaling invariant but the presence of a non-trivial dilaton profile breaks it. The field strength vanishes whereas the dilaton diverges, which goes logarithmically\footnote{
We leave the study of the dispersion relation obeyed by these solutions, along the lines of \cite{Karch:2008fa} and \cite{Brattan:2012nb}, for future research.}. For generic $\theta=\gamma(d-1)$ and $\delta=1$, the finite temperature  metric looks as
\be
ds^2_{d+1}=r^{-2\gamma}\left[-r^{2z}f(r)dt^2+r^2 dx^idx^j\delta_{ij}+\f{dr^2}{r^2f(r)}\right],\quad f(r)=1-\left(r_h/r\right)^{ z+(1-\gamma)(d-1)}
\ee

For $\delta=0,~ z=2$ and $\theta\neq 0$, the solution is very promising in $3+1$ dimensional spacetime, in the sense, that it gives not only the linear temperature dependence of the resistivity but it gives the logarithmic violation of the entanglement entropy. In which case, the geometry is conformal to $AdS_2\otimes R^{d-1}$ and the finite temperature geometry   reads as
\be
ds^2_{d+1}=R^2 u^{-2\gamma/z}\left[-u^{2}f(u)dt^2+ dx^idx^j\delta_{ij}+\f{du^2}{u^2f(u)}\right], \quad 
f(u)=1-\left(u_h/u\right)^{ \f{z+(1-\gamma)(d-1)}{z}}
\ee
where we have reinstated the size, $R$, of the  $AdS_2$.

Moving onto the solution at UV,  in the presence of the dilaton  there exists  an electrically charged black hole solution in any arbitrary spacetime dimensions, whose geometry reads as 

\be
ds^2_{d+1}=-r^2 f(r) dt^2+r^2 dx^idx^j\delta_{ij}+\f{dr^2}{r^{2(1-\delta_1)}f(r)}
\ee
and the form $f(r)$ is written in eq(\ref{f_dilaton_dbi}). The positivity of the specific heat at low charge density suggests to have a restriction on the parameter, $\delta_1 < 1/2$.   In $3+1$ dimensional spacetime, we found a 
 dyonic black hole solution.  In the limit of $\delta_1\ra 0$, we do reproduce an asymptotically charged $AdS_{d+1}$ spacetime. 
Interestingly, it is only the electrically charged black hole solution are same as that obtained for the BI black holes. 
%in \cite{Yamazaki:2001ue}, where the authors have found the solution only in $2+1$ dimensional spacetime.
These black holes, both with and without the  dilaton field, have some non-zero entropy even in the zero temperature limit.  We leave the detailed study of the thermodynamics for future investigations. Instead of considering the space filling D$d$-brane action, it would be interesting to study the lower dimensional brane action along the lines of \cite{Giddings:2001yu} with the massive embeddings \cite{Mateos:2006yd}, whose  solution as well as the rich  thermodynamics, we leave for  future research.

There exist one more avenue for future research that is to  find solution that  interpolates between the UV  solution eq(\ref{geometry_dilaton_dbi}) and IR solution eq(\ref{most_general_geometry_with_all_possibilities}) found in this work.\\

{\bf Acknowledgment}\\
It is a pleasure to thank Bum-Hoon Lee,  Shibaji Roy and  Tadashi Takayanagi
 for their useful comments, especially to Sean A. Hartnoll, Elias Kiritsis and the anonymous referee for several useful comments and suggestions. Thanks are 
 to  Saha institute of Nuclear Physics, Kolkata and  CQUeST, Sogang university, Seoul for their generous help and support at different stages of  this work. 

\section{Appendix A}

In this section, we shall generalize  the scaling behavior of  $AdS_2\otimes R^{d-1}$ geometry and the hyperscaling violating geometry as studied e.g., in \cite{Huijse:2011ef}. Essentially, we are combining these two different kind of spacetimes and obtain the most   
general scaling behavior of the spacetime that respects the rotational and translational symmetry. 

Let us write down the most general geometry based on these symmetries  but in a restricted sense.
Essentially,  the metric components follows a power law type behavior and for simplicity, we take the metric components to be functions of the radial coordinate only
\be
ds^2_D=R^2 r^{2a}\left[-r^{2b}dt^2+r^{2c}dx^2_i+\f{dr^2}{r^{2d}}\right],\quad i=1,2,\cdots,D-2.
\ee

For non-zero, $c$, we can re-write  the geometry by defining a new coordinate  via  $u=r^c$
\be
ds^2_D=R^2 u^{2a/c}\left[-u^{2b/c}dt^2+u^{2}dx^2_i+\f{du^2}{u^{2(c+d)/c}}\right],
\ee
where we have re-defined  $t,~x_i$ and  $R$ as well. Now, demand that under 
$u\ra \lambda^{\alpha} u$, time, spatial coordinates, $x_i$ and the metric scales as
\be
t\ra \lambda^z t,\quad x_i\ra  \lambda^{\delta} x_i,\quad ds\ra  \lambda^{\beta} ds.
\ee

In which case, the exponents $a/c,~b/c$ and $d/c$ are related to $\beta,~\delta$ and $z$ as
\be
a/c=-\f{1}{\alpha}(\delta+\alpha-\beta),\quad  b/c=\f{1}{\alpha}(\delta+\alpha-z),\quad \f{(c+d)}{c}=-\delta/\alpha
\ee
and the geometry becomes 
\be
ds^2_D=R^2 u^{-2\f{(\delta+\alpha-\beta)}{\alpha}}\left[-u^{\f{2(\delta+\alpha-z)}{\alpha}}dt^2+u^2 dx^2_i+du^2 u^{2\delta/\alpha} \right].
\ee

Note that $\alpha$ can not be set to zero, in fact, for any other value of it, we can redefine $\delta,~\beta$ and $z$ and bring the geometry to the following form
\be
ds^2_D=R^2 u^{-2(1+{\tilde\delta}-{\tilde\beta})}\left[-u^{2(1+{\tilde\delta}-{\tilde z})}dt^2+u^2 dx^2_i+du^2 u^{2{\tilde\delta}} \right]=R^2u^{2{\tilde\beta}}\left[-u^{-2{\tilde z}}dt^2+u^{-2{\tilde\delta}}dx^2_i+\f{du^2}{u^2} \right],
\ee
where ${\tilde\delta}=\delta/\alpha,~{\tilde\beta}=\beta/\alpha$ and ${\tilde z}=z/\alpha$. Since, we are interested to understand the scaling behavior of the geometry, it is better not to set $\alpha$ to zero. Now, it just follows that under

\be
u\ra \lambda u,\quad t\ra \lambda^{{\tilde z}} t,\quad x_i\ra \lambda^{{\tilde\delta}} x_i,\quad ds\ra \lambda^{{\tilde\beta}} ds.
\ee

Even though, we started out with a geometry which has non-linear scaling behavior of the radial coordinate, $u$, but after redefinition of the exponents, the radial coordinate has been set to scale linearly. Moreover, for ${\tilde \delta}\neq 0$, we can redefine the coordinate $u^{\tilde\delta}=1/\rho$ also $t,~x_i$ and $R$ to make the $x_i$\rq{}s  scale linearly in $\lambda$, i.e., $x_i\ra \lambda x_i$.
In fact, the hyperscaling violating  geometry as written down in \cite{Huijse:2011ef} has been re-written in  
\cite{Dong:2012se} for which  the radial coordinate scales linearly. So, we finally end up with two different choices of  ${\tilde \delta}$. Those are zero and unity.

Let us record the curvature invariants of eq(\ref{most_general_geometry_with_all_possibilities}) in $d=3$
\bea
{\rm Ricci~ scalar}&=&2r^{2\gamma}[z(3 \gamma-2\delta)-3(\gamma-\delta)^2-z^2],\nn
R^{MN}R_{MN}&=&r^{4\gamma}\Bigg[(z^2-z\gamma+2\delta(\delta-\gamma))^2+2(\gamma-\delta)^2(z-2\gamma+2\delta)^2+\nn&&
(z-\gamma)^2(z-2\gamma+2\delta)^2 \Bigg],\nn
R^{MNKL}R_{MNKL}&=&4r^{4\gamma}\Bigg[z^4-2z^3\gamma-4z\gamma(\gamma-\delta)^2+z^2(3\gamma^2-4\gamma\delta+2\delta^2)
+\nn &&(\gamma-\delta)^2(3\gamma^2-2\gamma\delta+3\delta^2) \Bigg]
\eea

These invariants suggests that for $\gamma<0$ the geometry is singular at IR, which  is the case for the scale symmetry violating solutions.

\section{Appendix B}

In this section, we shall find the exact solution to the Einstein-Hilbert- cosmological constant action along  with the BI action. In particular, with an electric and constant  magnetic field.
On finding the solution we shall see the difference of this solution with that found in section 5.1, namely, the Einstein-Hilbert- cosmological constant  action with the  DBI term. 
%In this case there exists a constant magnetic field along with the electric field. 
The BI action is described as 
\be
S_{BI}=-T\int \sqrt{-g}\sqrt{1+\alpha F^{MN}F_{MN}}.
\ee
For small value of $\alpha$ this action reduces to the Maxwellian action. From the study of the Einstein-Maxwell-cosmological constant system in $3+1$ dimensional spacetime, it is known that given an electrically charged solution with charge density $\rho$, we can obtain a dyonic solution by doing the following substitution: $\rho^2\ra \rho^2+B^2$, where $B$ is the constant magnetic field. However, this simple substitution does not work for the Einstein-Hilbert-cosmological constant-BI system, which we demonstrate below. 

The full action is 
\be
S=\f{1}{2\kappa^2}\int d^{3+1} x \sqrt{-g}\bigg(R-2\Lambda-T\sqrt{1+\alpha F^{MN}F_{MN}}\bigg).
\ee

The equation of motion for the metric and the gauge field  are
\be
R_{MN}-\Lambda g_{MN}-\f{T g_{MN}}{2\sqrt{1+\alpha F^2}}-\f{T\alpha g^{KL}F_{MK}F_{NL}}{\sqrt{1+\alpha F^2}}=0,\quad 
\p_M\left(\f{\sqrt{-g} F^{MN}}{\sqrt{1+\alpha F^2}} \right)=0,
\ee
where we use a short hand notation $F^2= F^{MN}F_{MN}$. Let us consider the following ansatz for the gauge field and the metric
\be
ds^2=-g_{tt}(r)dt^2+g_{xx}(r)(dx^2+dy^2)+g_{rr}(r)dr^2\quad F=A\rq{}_t(r) dr\w dt+B dx\w dy.
\ee

Substituting such a choice of the gauge field strength  into the equation of motion gives 
\be
A\rq{}_t(r)=\rho \f{\sqrt{g_{tt}g_{rr}(g^2_{xx}+2\alpha B^2)}}{g_{xx}(g^2_{xx}+2\alpha \rho^2)},
\ee
where $\rho$ is the constant of the integration and we interpret it as the charge density. The  equations of motion of the metric components are
\bea
&&R_{tt}+\Lambda g_{tt}+\f{T}{2} \f{g_{tt}(g^4_{xx}-4\alpha^2\rho^2 B^2)}{g^2_{xx}\sqrt{(g^2_{xx}+2\alpha B^2)(g^2_{xx}+2\alpha \rho^2)}}=0,\nn
&&R_{xx}-\Lambda g_{xx}-\f{T}{2 g_{xx}}\sqrt{(g^2_{xx}+2\alpha B^2)(g^2_{xx}+2\alpha \rho^2)}=0,\nn 
&&R_{rr}-\Lambda g_{rr}-\f{T}{2} \f{g_{rr}(g^4_{xx}-4\alpha^2\rho^2 B^2)}{g^2_{xx}\sqrt{(g^2_{xx}+2\alpha B^2)(g^2_{xx}+2\alpha \rho^2)}}=0.
\eea

Let us demand that the asymptotic solution to the above equations of motion is AdS, which means we can set the metric as
\be
ds^2=\left(\f{r}{R}\right)^2\left[-f(r)dt^2+dx^2+dy^2\right]+\f{R^2 dr^2}{r^2 f(r)}.
\ee

Upon substituting it into the above equations of motion, one can easily convince that the $g_{tt}$ and the $g_{rr}$ component of the equations of motion are not independent of each other and it reads as
\be
r^2f\rq{}\rq{}+6 rf\rq{}+6f+2\Lambda R^2+T R^2\f{(r^8-4B^2R^8\alpha^2\rho^2)}{r^4\sqrt{(r^4+2\alpha R^4 B^2)(r^4+2\alpha R^4\rho^2)}}=0
\ee

In fact the  equation of motion associated to $g_{xx}$ gives
\be
rf\rq{}+3f+\Lambda R^2+T R^2\f{\sqrt{(r^4+2\alpha R^4 B^2)(r^4+2\alpha R^4\rho^2)}}{2r^4}=0.
\ee

It is easy to convince that these two differential equations are not independent of each other. Moreover, by comparing this differential equation for $f(r)$ with that of the DBI case as in eq(\ref{diff_dyonic_dbi}), we can easily convince that they are not same 
%the difference of the  Einstein-BI-cosmological constant system with that of the  Einstein-DBI-cosmological constant system 
as far as the dyonic solution is concerned. As a result,  it simply follows that given  an electrically charged solution of the 
 Einstein-Hilbert-cosmological constant-BI system  with charge density 
$\rho$  can not give a dyonic solution  by the simple substitution formula: $\rho^2\ra \rho^2+B^2$.

\end{document}